\documentclass[%
 prl, twocolumn,
 amsmath,amssymb,
 aps, superscriptaddress]{revtex4-2}

\usepackage{graphicx}
\usepackage{dcolumn}
\usepackage{bm}
\usepackage{dsfont}
\usepackage{xcolor}
\usepackage{mathrsfs}
\newcommand{\bs}{\boldsymbol}
\usepackage[colorlinks=true,citecolor=blue]{hyperref}

\begin{document}

\title{Topological charge, spin and heat transistor}

\author{V. Fern\'andez Becerra}
 \email{becerra@magtop.ifpan.edu.pl}
\affiliation{International Research Centre MagTop, Institute of Physics, Polish Academy of Sciences, Aleja Lotnikow 32/46, PL-02668 Warsaw, Poland}
\author{Mircea Trif}
\affiliation{International Research Centre MagTop, Institute of Physics, Polish Academy of Sciences, Aleja Lotnikow 32/46, PL-02668 Warsaw, Poland}
\author{Timo Hyart} 
\affiliation{International Research Centre MagTop, Institute of Physics, Polish Academy of Sciences, Aleja Lotnikow 32/46, PL-02668 Warsaw, Poland}
\affiliation{Department of Applied Physics, Aalto University, 00076 Aalto, Espoo, Finland}

\date{\today}

\begin{abstract}
Spin pumping consists in the injection of spin currents into a non-magnetic material due to the precession of an adjacent ferromagnet. In addition to the pumping of spin the precession always leads to pumping of heat, but in the presence of spin-orbital entanglement it also leads to a charge current. We investigate the pumping of charge, spin and heat in a device where a superconductor and a quantum spin Hall insulator are in proximity contact with a ferromagnetic insulator. We show that the device supports two robust operation regimes arising from topological effects. In one regime, the pumped charge, spin and heat are quantized and related to each other due to a topological winding number of the reflection coefficient in the scattering matrix formalism -- translating to a Chern number in the case of Hamiltonian formalism. In the second regime, a Majorana zero mode switches off the pumping of currents owing to the topologically protected perfect Andreev reflection. We show that the interplay of these two topological effects can be utilized so that the device operates as a robust charge, spin and heat transistor. 
\end{abstract}

\maketitle

{\it Introduction.}$-$ Transistors are a celebrated example of a basic research discovery resulting in an enormous societal impact. They are the building blocks of the modern digital technology revolution owing to their ability to manipulate electrical currents with exponential dependencies on the control parameters \cite{electrobook}. Motivated by this success story, enormous amount of research efforts have been devoted to enhancing the functionalities of the next generation of the nanoelectronic devices by exploiting the various ways to manipulate the electrical currents on the level of single electrons \cite{KastnerRMP92, Egger12, Pekola13}, the spin degree of freedom \cite{Datta-Das, Spbattery, Spump_theo, TserkovnyakRMP2005, Fert08, Grunberg08, Qquanti_Zhang, qtizedspin, Wunderlich1801, Fu14, Spump+TIone, Spump+TItwo, SinovaRMP2015, Spump+TIthree, spintroRev}, the thermal properties of mesoscopic structures \cite{Heikkila06, Pekola15, Fornieri2017, heatpump}, as well as the mutual coupling of the charge, spin and energy modes \cite{Heikkila18, Virtanen20}.

\begin{figure}[t]
\includegraphics[width=0.9\linewidth]{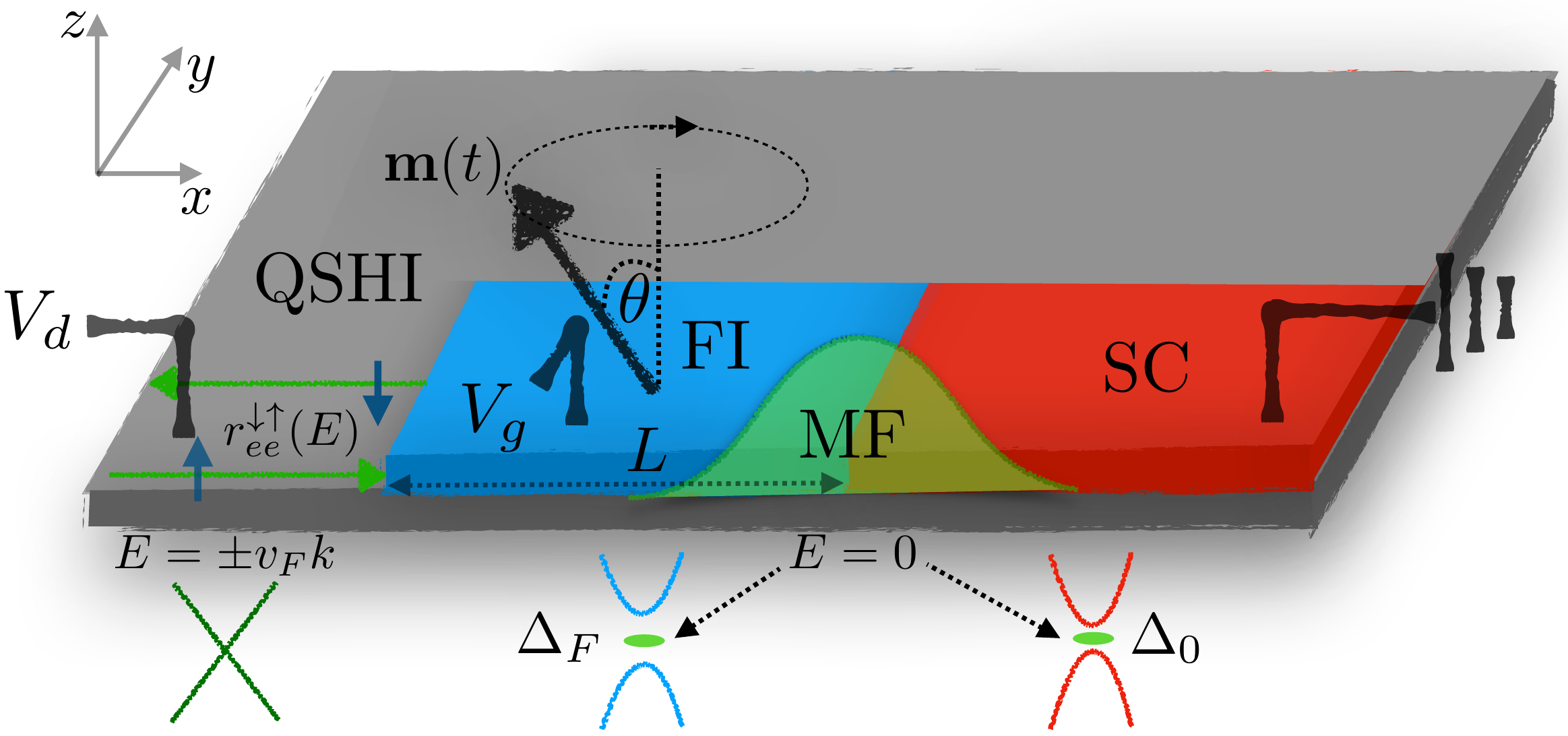}
\caption{Sketch of the proposed device consisting of a QSHI in proximity contact with FI of length $L$ (induced energy gap $\Delta_{F}=|\mathbf{m}| \sin\theta$) and SC (induced energy gap $\Delta_0$). The monodomain magnetization ${\bs m}(t)$ precesses at an angle $\theta$ around the axis perpendicular to the QSHI driving charge, spin and heat currents to the drain (left), which can be controlled with potential at the FI region $V_g$, the precession angle $\theta$, temperature and drain voltage $V_d$. The system harbors a zero-energy MF at the FI-SC interface that affects the pumped currents via the scattering coefficient $r_{ee}^{\downarrow\uparrow}(E)$.}  
\label{fig1}
\end{figure}

Topological materials are the golden standard for future electronic, spintronic and heattronic devices as the corresponding transport modes are intrinsically linked to each other in these systems \cite{RevKane,RevZhang}. This is best exemplified in the case of two-dimensional quantum spin Hall insulators (QSHI), which support one-dimensional helical edge modes, so that the electrons moving right and left carry opposite spins \cite{Kane-mele,BHZ}. The QSHI states have been observed in various materials \cite{QSH_HgTe, Du15, Wu18}, but their potential for device applications is still waiting to be realized. Nevertheless, from the previously explored topological phenomena, quantum Hall effect and ac Josephson effect, we know that topological effects are well-suited for metrology applications. The quantum Hall has been widely used as a resistance standard \cite{KlitzingPTRS2011} and ac Josephson effect as a voltage standard \cite{HamiltonAIP2000}.

In this Letter, we consider a system where the QSHI edge is placed in proximity to a ferromagnetic insulator (FI) and a superconductor (SC) to realize Majorana zero-energy mode (MF) \cite{Kane_prox, fu2009josephson, Beenakker-review} (see Fig.~\ref{fig1}). The technology for building this setup has already been developed motivated by the prospects of utilising MFs as a staple ingredient for topological quantum computers \cite{yacobyNP2014, PribiagNatNano2015, YuNanoLett2020, Vaitieknas2020ZerofieldTS, Lutchyn2018}. We explore the potential of this setup in a completely different context utilising the perfect topologically protected Andreev reflection (AR) enabled by the MF \cite{LawPRL2009, Beenakker-review} but not exploiting the existence of a nonlocal quantum degrees of freedom used in topological quantum computers \cite{RevMajoforQcomp}. Namely, we study the charge, spin and heat pumping in this system in the presence of precessing magnetization (see Fig.~\ref{fig1}). The spin pumping is a scrutinized method to generate spin currents in magnetic heterostructures \cite{Spbattery, Spump_theo, TserkovnyakRMP2005, SinovaRMP2015} and forms the basis for many contemporary spintronic applications \cite{Hoffman-review}, but in the case of QSHIs there exists a unique property that both the charge and spin pumping are quantized and related to each other \cite{Qquanti_Zhang, qtizedspin}. We show that also the heat current is quantized and elaborate the origin of these quantizations by showing that they result from a topological winding number of reflection coefficient in the scattering matrix formalism and Chern number in the Hamiltonian formalism -- resembling the quantized topological pumps proposed in other contexts \cite{Thouless83, AvronRMP, AvronPRB, Nazarov2016}. Then we show that the key advantage of our QSHI-FI-SC heterostructure in comparison to the earlier proposals is that due to the presence of MF there exists also another topological operation regime, where the MF switches off the pumping of currents owing to the perfect AR. We show that it is possible to use external control parameters to tune between these operation regimes so that the device operates as a robust charge, spin and heat transistor with exponential sensitivity on the applied gate voltage and precession angle.  Moreover, two perfectly quantized limits allow to build standards for the spin and heat pumping with the help of accurate measurement of the pumped electric charge.

{\it Theoretical approach.}$-$ To describe the system, consisting of the edge states of 2D QSHI in proximity contact with FI and SC as illustrated in Fig.~\ref{fig1}, we consider a time-dependent Bogoliubov-de Gennes (BdG)  Hamiltonian
\begin{align}
    \mathcal{H}_{BdG}(t)\!=\![v_Fp\,\sigma_z\!-\!\mu(x)]\tau_z\!+\!{\bs m}(x,t)\cdot{\bs \sigma}\!+\!\Delta(x)\tau_x\,,
\end{align}
where ${\bs \sigma}=(\sigma_x,\sigma_y,\sigma_z)$ and ${\bs \tau}=(\tau_x,\tau_y,\tau_z)$ are Pauli matrices that act in the spin and Nambu space respectively, $v_{F}$ is the Fermi velocity, $p=-i\hbar\partial_x$ is the momentum operator along the edge ($x$-direction),   ${\bs m}(x,t)$ is the time-dependent magnetization in the FI (which includes the exchange coupling strength between the two materials), $\Delta(x)$ is the induced superconducting order parameter and $\mu(x)$ is the chemical potential. We assume that $\Delta(x)=\Delta_0$ is constant  over the region occupied by the SC. The magnetization of the FI island is parametrized as ${\bs m}(x,t)=m_0(x)[\sin{\theta(t)}\cos\phi(t),\sin{\theta(t)} \sin{\phi(t)},\cos{\theta(t)}]$, where $m_0(x)=m_0$ under the FI region (uniform precession). Moreover, we consider periodic driving such that ${\bs m}(x,t+\mathcal{T})={\bs m}(x,t)$, with $\mathcal{T}=2\pi/\omega$ being the precession period and $\omega$ the precession frequency. Throughout the text we assume that the temperature is much smaller than the critical temperatures of the superconductivity and magnetism, so that the temperature dependence of $\Delta_0$ and $m_0$ can be neglected.

The dynamics of the magnetization in the FI results in pumping of charge, spin  and  heat into the left lead. The pumped charge over one cycle in the adiabatic limit can be calculated  from the expression \cite{MoskaletsBook, Blaauboer} 
\begin{equation}
Q_e =-\frac{e}{4\pi} \int\!dE \Bigl(\frac{\partial f}{\partial E}\Bigr)\!
\int_0^{\mathcal{T}} \!dt\, \mathrm{Im}\Bigl\{\mathrm{Tr}\Bigr[ \mathcal{S}^{\dagger} \tau_z \frac{\partial \mathcal{S}}{\partial t} \Bigr]\Bigr\}\,,
\label{qqBPT}
\end{equation}
where $f(E)$ is the Fermi distribution function, and $\mathcal{S}(E, t)\equiv \mathcal{S}(E,\theta(t), \phi(t))$ is the instantaneous scattering matrix pertaining to a normal metal-FI-SC junction. This  can be casted in the form 
\begin{equation}
    \mathcal{S}(E,\theta, \phi) \equiv \left(
    \begin{array}{cc}
    \mathcal{S}^{ee}(E,\theta, \phi) & \mathcal{S}^{eh}(E, \theta, \phi) \\
    \mathcal{S}^{he}(E,\theta, \phi) & \mathcal{S}^{hh}(E,\theta, \phi)
    \end{array}
    \right)\,,
    \label{Smat}
\end{equation}
accounting for both the normal (ee) and Andreev (eh) processes, so that each of these components is a matrix describing the spin-dependent scattering. The pumping of spin ${\bs S}\,$  can be found analogously by using substitutions $e\rightarrow \hbar/2$ and $\tau_z\rightarrow{\bs \sigma}$ in Eq.~(\ref{qqBPT}). We consider only  $S_z$ component since it is the only spin-component conserved in the left lead. Finally, the heat $Q_E$ injected in the left lead  is  obtained from the expression 
\begin{equation}
Q_E =-\frac{\hbar}{8\pi} \int\!dE \Bigl(\frac{\partial f}{\partial E}\Bigr)\!
\int_0^{\mathcal{T}} \!dt\, \mathrm{Tr}\Bigr[\frac{\partial \mathcal{S}}{\partial t}\frac{\partial \mathcal{S}^{\dagger}}{\partial t}\Bigr]\,.
\label{heatBPT}
\end{equation}
Here, we have neglected the possible heat losses to the substrate. The spin-momentum locking in the QSHI edges limits the scattering matrix elements  so that the only non-zero reflection coefficients are  $r_{ee (hh)}^{\downarrow\uparrow}(E,\theta, \phi)$  and  $r_{he(eh)}^{\downarrow\uparrow}(E,\theta)$. Here,  $r_{ee(hh)}^{\downarrow\uparrow}(E,\theta, \phi)$ describes the reflection amplitude for an electron (hole) with spin $\uparrow$ injected from the QSHI at energy $E$ to be reflected back to QSHI as electron (hole) with spin $\downarrow$ because of the FI and the SC. Similarly, $r_{he(eh)}^{\downarrow\uparrow}(E,\theta)$ describe the amplitudes for the AR processes, where electron (hole) is reflected back as a hole (electron). Each reflection coefficient accounts for all the possible scattering paths, including the effect of the MF at the FI-SC interface, and the reflection coefficients satisfy $|r_{ee (hh)}^{\downarrow\uparrow}(E,\theta, \phi)|^2+|r_{he(eh)}^{\downarrow\uparrow}(E,\theta)|^2=1$.

The only $\phi$-dependent coefficients satisfy \cite{supplementary}
\begin{equation}
r_{ee}^{\downarrow\uparrow}(E, \theta, \phi)=r_0 (E, \theta) e^{i \phi}, \ r_{hh}^{\downarrow\uparrow}(E)=-[r_{ee}^{\downarrow\uparrow}(-E)]^*,
\end{equation}
and the magnitude of $r_{ee}^{\downarrow\uparrow}(E, \theta, \phi)$ is suppressed at low-energies due to the topologically protected perfect AR $|r_{he}^{\downarrow\uparrow}(E=0,\theta)|=1$, so that for $E \ll m_0, \Delta_0$ it can be approximated as \cite{supplementary}
\begin{equation}
    |r_{0}(E, \theta)|^2 \approx  \frac{E^2/\Gamma^2}{1 + E^2/\Gamma^2} ,
    \label{ree_MAIN}
\end{equation}
where
\begin{equation}
    \Gamma = 2\Delta_0\left(\frac{\xi_F(0,\theta)}{\xi_F(V_g,\theta)}\right)^2
    \frac{\xi_S}{\xi_F(V_g,\theta)+\xi_S}
    e^{-2L/\xi_{F}(V_g,\theta)}\, 
\label{formGamma}
\end{equation}
is the Majorana linewidth (for which $|r_{ee}^{\downarrow\uparrow}(E=\Gamma)|^2=|r_{he}^{\downarrow\uparrow}(E)|^2=1/2$) expressed in terms of the coherence lengths
$\xi_F(V_g,\theta)=\hbar v_F/{\sqrt{m_0^2\sin^2\theta-(eV_g)^2}}$ and $\xi_S=\hbar v_F/\Delta_0$ pertaining to the ferromagnet and superconductor, respectively. Here, we have denoted the chemical potential in the FI region as $\mu_{_{FI}}=eV_g$ to indicate that it can be controlled with the gate voltage and $L$ is the length of the FI region (Fig.~\ref{fig1}).
In analytic calculations we use the approximation (\ref{ree_MAIN}), but  numerical calculations are done using the full expression for $r_{ee}^{\downarrow\uparrow}(E, \theta, \phi)$  \cite{supplementary}. In the following we utilize the topological protection of the MFs, which ensures that $\Gamma$ depends exponentially on the parameters $V_g$ and $\theta$ [Eq.~(\ref{formGamma})].  This dependence only breaks down if  $e V_g\rightarrow m_0\sin\theta$, at which point the electrons under the FI become gapless and the adiabatic scattering matrix approximation is no longer valid.

For simplicity, in the following we assume $\phi(t)=\omega t$ and $\theta(t)\equiv\theta$ is  constant. However, many of our results can be generalized for  arbitrary trajectory in $(\theta(t), \phi(t))$-space \cite{supplementary}. Interestingly, the special form of the scattering matrix for the  combined system leads to the charge, spin and heat pumped over a cycle to be determined by a single dimensionless charge $\mathcal{Q}$: 
\begin{equation}
Q_e = e\mathcal{Q}, \quad S_z = -\frac{\hbar}{2}\mathcal{Q}, \quad Q_E = \frac{\hbar\omega}{2}\mathcal{Q}\,.
\label{universal}
\end{equation}
In the adiabatic limit
\begin{equation}
\mathcal{Q} =-\frac{1}{2\pi}\! \int\!dE \Bigl(\frac{\partial f(E)}{\partial E}\Bigr)
\int_0^{2\pi} \!d\phi\,\bigl |r_{ee}^{\downarrow\uparrow}(E, \theta, \phi)\bigr|^2\, \label{Qadiabatic-main}
\end{equation}
depends on the applied drain voltage $V_d$ and temperature $k_BT$ ($k_B$ is the Boltzmann constant) via  $f(E)$, and gate voltage $V_g$ and rotation angle $\theta$ via  $|r_{ee}^{\downarrow\uparrow}(E, \theta, \phi)\bigr|^2$.  In the special case of constant $\theta$ we can also go the rotating frame and calculate the frequency dependence of $\mathcal{Q}$ \cite{supplementary}. In the limit  $T=0$ and $V_d=0$, we obtain  
\begin{equation}
 \mathcal{Q}= \bigg[1-\frac{2 \Gamma}{\hbar \omega}\arctan\bigg(\frac{\hbar \omega}{2 \Gamma}\bigg) \bigg]\,.
\end{equation}
In a continuous operation of the device the pumped charge, spin, and heat currents flowing into the drain can be written, respectively, as $\langle I_e\rangle= \omega\,Q_e/2\pi$, $\langle I_s\rangle=\omega\,S_z/2\pi$, and $\langle I_E\rangle= \omega\,Q_E/2\pi$.
If drain voltage is applied ($V_d \ne 0$) there exist also dc current contributions which do not depend on the magnetization dynamics. In this case the pumped currents can be obtained by measuring the currents in the presence and absence of the magnetization dynamics \cite{supplementary}.

{\it Topological effects.}$-$ As shown above, all quantities of interest are determined by $\mathcal{Q}$. Before analyzing  $\mathcal{Q}$ in detail, we first highlight the robust topological features of this quantity. Namely, we find that  $\mathcal{Q}=0$ for $eV_d, k_BT, \hbar \omega \ll\Gamma$ owing to the perfect topological AR caused by the MF. On the other hand, $\mathcal{Q}=1$ if $eV_d \gg \Gamma$, or $k_BT \gg \Gamma$ or $\hbar \omega \gg \Gamma$.  The latter quantization appears due to the topological winding of the phase of  $r_{ee}^{\downarrow\uparrow}(E, \theta, \phi)=r_0 (E, \theta) e^{i \phi}$ when the above conditions guarantee that the magnitude satisfies $|r_{ee}^{\downarrow\uparrow}(E, \theta, \phi)\bigr|=1$. The quantization of the pumped charge to $\mathcal{Q}=1$ can also be understood as Thouless pumping arising due to a Chern number associated with the pumping cycle in the case of the Hamiltonian formalism \cite{supplementary}.   The quantized charge and spin pumping in these limits are robust topological results which are independent of the details of the magnetization trajectory $(\theta(t), \phi(t))$ \cite{supplementary}. The topological protection guarantees that deviations of  $\mathcal{Q}$ on the quantized values depend on an exponential way on the parameters $V_g$ and $\theta$.

On the other hand, the quantization of the heat $Q_E$ can be related to the mesoscopic charge relaxation in quantum capacitors \cite{ThomasPLA93,NiggPRL06}. There, the charge relaxation resistance associated to a single  conduction channel coupled to the mesoscopic capacitor is $R_q=h/2e^2$ leading to $\langle I_E\rangle=R_q\langle I_e^2\rangle$. For circular precession, we find $\langle I_e^2\rangle=(\langle I_e\rangle)^2$, and hence the quantization of the charge $Q_e$ engenders quantization of $Q_E$. Nevertheless, contrary to the mesoscopic capacitors, here the quantization stems from the topological gap of the system and not from  the discreteness of the energy levels. 

\begin{figure}[t!]
\centering
\includegraphics[width=\linewidth]{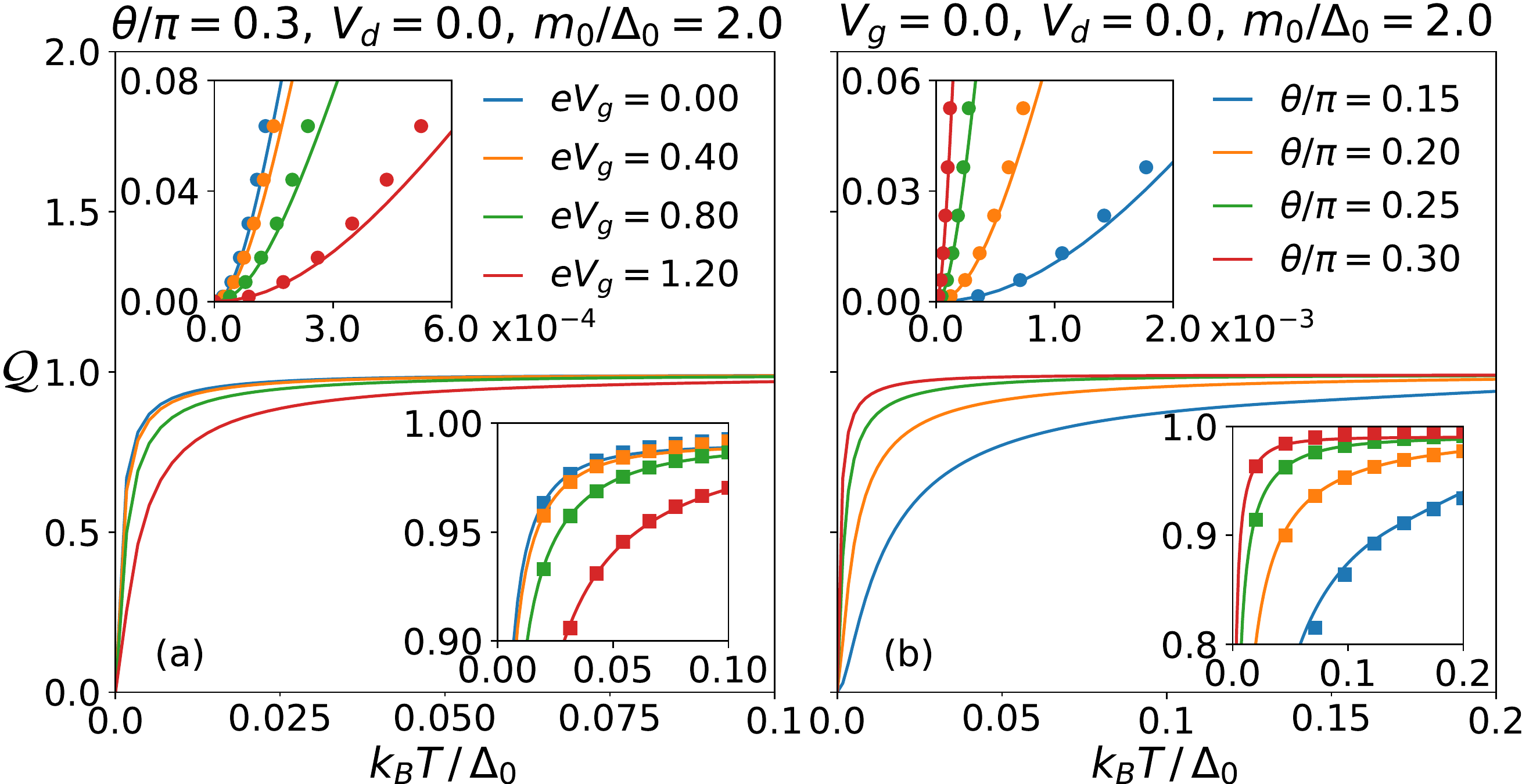}
\caption{The dimensionless charge $\mathcal{Q}$ in the adiabatic limit [Eq.~(\ref{Qadiabatic-main})]  as a function of $T$ for various values of (a) gate voltages $eV_g$ (in units $\Delta_0$) and (b) angles $\theta$. The insets are zooms into the suppression ($k_B T \ll \Gamma$, $\mathcal{Q} \approx 0$) and quantization ($k_B T \gg \Gamma$, $\mathcal{Q} \approx 1$) regimes. The asymptotic expressions of $\mathcal{Q}$ [Eq.~(\ref{asymptotic-T})] are shown in the suppression (circles) and  quantization (squares) regimes.  The other  parameters are  $L=400$ nm, $\Delta_0 = 1$ meV, $m_0/\Delta_0 = 2$ and $v_F = 2.7\times 10^5$ m/s. }
\label{IvsT}
\end{figure}

{\it Transistor behaviour.}$-$ The detailed analysis of the transistor characteristics can be based on the interplay of the topological effects discussed above. We start by discussing the  characteristics in the absence of drain voltage $V_d=0$ because in this case the currents are caused purely by the magnetization dynamics.

In Fig.~\ref{IvsT},  we show the dimensionless charge $\mathcal{Q}$ in the adiabatic limit [Eq.~(\ref{Qadiabatic-main})] as a function of temperature $T$  for different values of $V_g$ and $\theta$. The asymptotic expressions of $\mathcal{Q}$ at low and high temperatures are \cite{supplementary}
\begin{align}
\mathcal{Q} \approx \left\{
\begin{array}{cc}
\displaystyle{\frac{\pi^2}{3}\left(\frac{k_BT}{\Gamma}\right)^2}
\,, & k_B T \ll \Gamma\, \\\\
1- \displaystyle{\frac{\pi}{4}\left(\frac{\Gamma}{k_BT}\right)
}\,, & k_B T \gg \Gamma\,
\end{array}
\right.\,. \label{asymptotic-T}
\end{align}
Due to the topological protection of MFs the Majorana linewidth $\Gamma$ depends exponentially on the control parameters $V_g$ and $\theta$ [Eq.~(\ref{formGamma})]. Therefore, it is easy to see from Eq.~(\ref{asymptotic-T}) and Fig.~\ref{IvsT} that the pumping can be efficiently turned on and off ensuring the device can operate as a transistor. Moreover, two perfectly quantized and extremely robust topological limits allow to define standards for charge, spin and heat pumping.

\begin{figure}[t!]
\centering
\includegraphics[width=\linewidth]{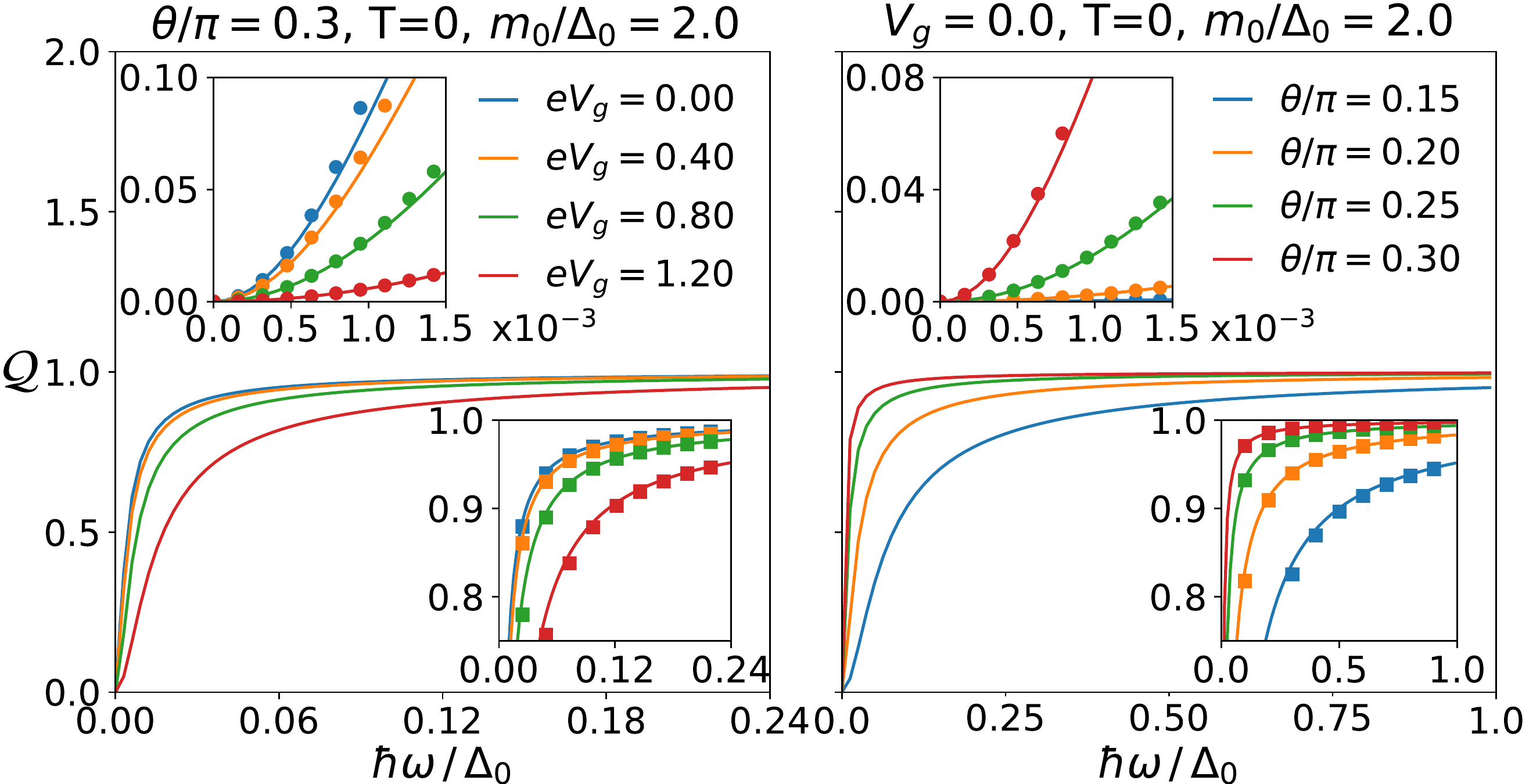}
\caption{The dimensionless charge $\mathcal{Q}$  as a function of $\omega$ for $V_d= T=0$ and various values of (a) $e V_g$ (in units of $\Delta_0$) and (b) $\theta$. The insets are zooms into the suppression ($\hbar \omega \ll \Gamma$, $\mathcal{Q} \approx 0$) and quantization ($\hbar \omega \gg \Gamma$, $\mathcal{Q} \approx 1$) regimes. The asymptotic expressions of $\mathcal{Q}$ [Eq.~(\ref{asymptotic-omega})] are shown in the suppression (circles) and  quantization (squares) regimes. The other parameters are same as in Fig.~\ref{IvsT}. }
\label{Ivsw}
\end{figure}

The operation of the transistor can also be controlled with the frequency. In Fig.~\ref{Ivsw}, we show $\mathcal{Q}$ as a function of $\omega$. Again, we have two operation regimes, and the corresponding asymptotic expressions of $\mathcal{Q}$ are
\begin{equation}
\mathcal{Q} = \begin{cases} 
\displaystyle{\frac{1}{12} \left( \frac{\hbar\omega}{\Gamma} \right)^2}\,, & |\hbar \omega| \ll \Gamma \\ \\
\displaystyle{1- \frac{\pi \Gamma}{ \hbar |\omega|}}\,, & |\hbar \omega| \gg \Gamma\,
\end{cases}
\,, \label{asymptotic-omega}
\end{equation}
demonstrating that pumping can again be  turned on and off with the control parameters $V_g$ and $\theta$.

\begin{figure}[t!]
\includegraphics[width=\linewidth]{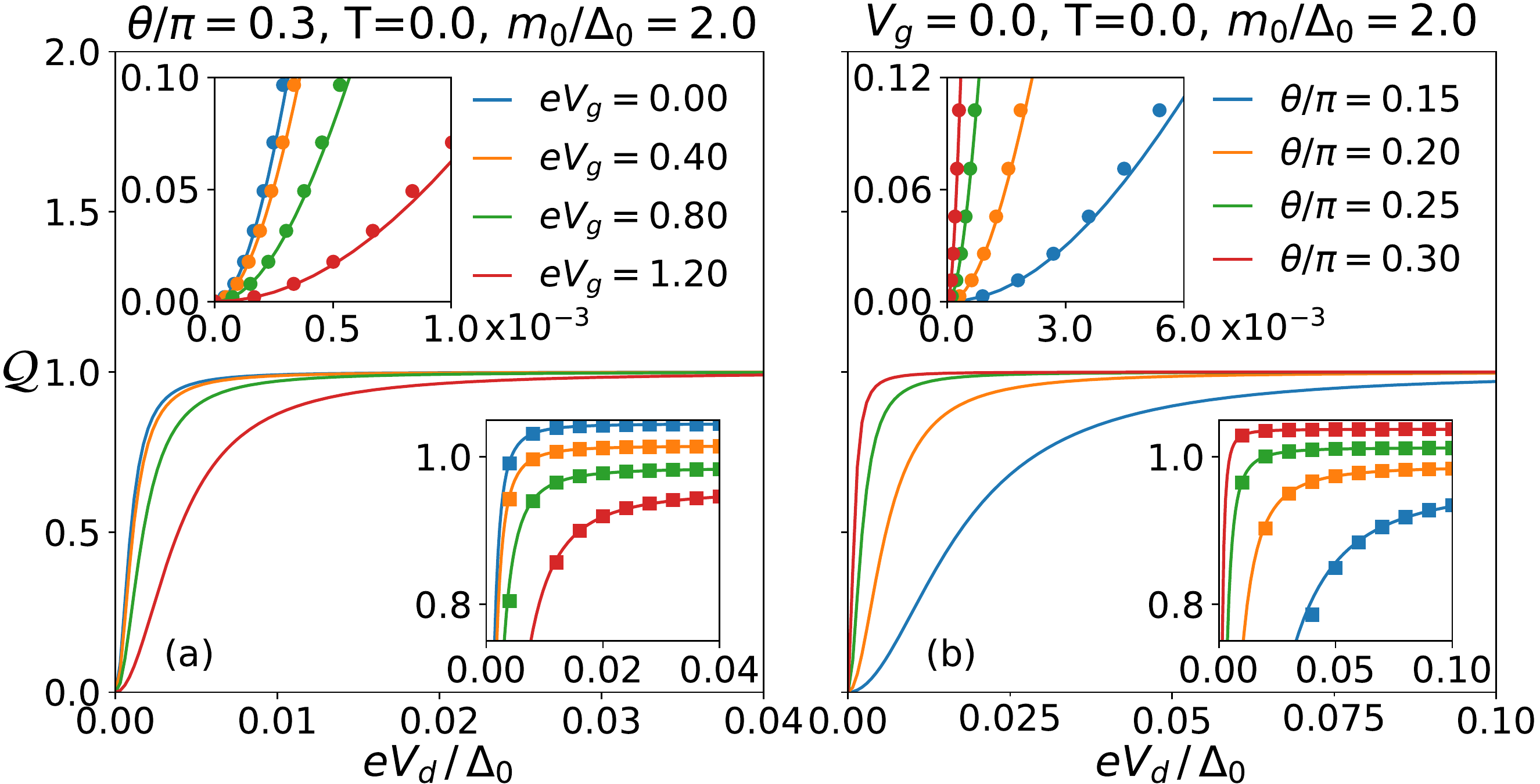}
\caption{The dimensionless charge $\mathcal{Q}$  as a function of drain voltage $V_d$ for various values of (a) gate voltages $V_g$ and (b) angles $\theta$. Two different limits for $\mathcal{Q}$ are evident in the plots. The insets are zooms into the suppression (low $V_d$) and quantization (high $V_d$) regimes, respectively. The asymptotic limits of the pumped charge are also shown in the suppression (circles) and  quantization (squares) regimes. The charges in the bottom insets have been shifted evenly for the sake of visualization.}
\label{IvsVd}
\end{figure}

Finally, in Fig.~\ref{IvsVd} we show the dimensionless charge $\mathcal{Q}$ in the adiabatic limit [Eq.~(\ref{Qadiabatic-main})] as a function of $V_d$. Similar robust switching behavior from $\mathcal{Q}=0$ to $\mathcal{Q}=1$ is obtained again with the asymptotic expressions
\begin{equation}
    \mathcal{Q} \approx \left\{
    \begin{array}{lc}
     \displaystyle{\left(\frac{eV_d}{\Gamma}\right)^2}\,,     & \quad eV_d\ll \Gamma     \\\\
     \displaystyle{1-\left(\frac{\Gamma}{eV_d}\right)^2}\,,   & \quad eV_d\gg \Gamma     
    \end{array}
    \right.\,,
\label{limits}    
\end{equation}
indicating that $\mathcal{Q}$ depends exponentially on $V_g$ and $\theta$. We empasize that for $V_d \ne 0$ also dc currents are present and the expressions (\ref{limits}) and Fig.~\ref{IvsVd} only describe the contribution coming from the magnetization dynamics.

In addition to the low-energy continuum model we have checked all features of the transistor behaviour using a Kwant software package \cite{Groth_2014} implementation of the 2D  quantum transport setup shown in Fig.~\ref{fig1}, including realistic disorder potential. The results from the two methods show excellent agreement \cite{supplementary}, solidifying the universality and robustness of the topological transistor. While the results have been obtained assuming circular precession of the FI magnetization around the $z$ axis, the quantization of the charge $Q_e$ and spin $S_z$ remain the same in the two topological limits, as long as the magnetization vector  encloses the $z$ axis during the precession.  On the other hand, the heat $Q_E$ will deviate from the universal expression in Eq.~(\ref{universal}) since in this case $\langle I_e^2\rangle\neq(\langle I_e\rangle)^2$ \cite{supplementary}.

{\it Conclusions and outlook} $-$ We have described the operation principles of a robust charge, spin and heat transistor consisting of a QSHI proximity coupled to FI and SC.  The device supports two robust operation regimes arising from topological effects. In the suppression regime at low energies the pumping is switched off due to the perfect AR of the electrons impinging on the MF hosted in the device.  Since the perfect AR is topologically protected, this suppression is not affected by disorder and other imperfections of the device. At high energies the pumped charge is quantized due to the topological winding number associated with the scattering matrix (or Thouless pumping). Therefore, the operation in this regime is also intrinsically robust against imperfections. The operation frequencies in our analysis are limited by the energy gap of the system $\hbar\omega\ll\Delta_0, m_0 \sin \theta$. Thus, the device can be operated at gigahertz frequencies, which is the typical frequency range of the spin pumping experiments.   Our device is scalable, as it is possible to pattern 2D QSHI with FI and SC arrays by various depositions methods.

\begin{acknowledgements}
{\em Acknowledgements.}
The work is supported by the Foundation for Polish Science through the IRA Programme
co-financed by EU within SG OP. Numerical simulations were carried out with the support of the Interdisciplinary Centre for Mathematical and Computational Modelling (ICM), University of Warsaw, under grant no G78-13.
\end{acknowledgements}

\bibliography{mainrefbiblio}

\end{document}


\title{Supplementary Material for ``Topological charge, spin and heat transistor"}

\author{V. Fern\'andez Becerra}
\affiliation{International Research Centre MagTop, Institute of Physics, Polish Academy of Sciences, Aleja Lotnikow 32/46, PL-02668 Warsaw, Poland}
\author{Mircea Trif}
\affiliation{International Research Centre MagTop, Institute of Physics, Polish Academy of Sciences, Aleja Lotnikow 32/46, PL-02668 Warsaw, Poland}
\author{Timo Hyart} 
\affiliation{International Research Centre MagTop, Institute of Physics, Polish Academy of Sciences, Aleja Lotnikow 32/46, PL-02668 Warsaw, Poland}
\affiliation{Department of Applied Physics, Aalto University, 00076 Aalto, Espoo, Finland}

\date{\today}

\maketitle

\appendix

\section{\label{sec:scattmatrixnormal} Pumped currents  in the absence of superconductivity}

Before studying the effect of superconductivity on the charge, spin and heat pumping we establish the properties of the pumping in the normal state. This analysis is also relevant for understanding the topological nature of our device at high energies where the presence of superconductivity introduces only exponentially small corrections to the values of the pumped charge, spin and heat. For this purpose we start by considering Hamiltonian  
\begin{align}
    H(t)=v_Fp\,\sigma_z-\mu(x)+{\bs m}(x,t)\cdot{\bs \sigma}\,. \label{Normal-supp}
\end{align}
where  ${\bs m}(x,t)=m_0(x)[\sin{\theta(t)}\cos\phi(t),\sin{\theta(t)} \sin{\phi(t)},\cos{\theta(t)}]$. Moreover, we consider a closed trajectory of the magnetization satisfying ${\bs m}(x,\mathcal{T})={\bs m}(x,0)$.

\subsection{Quantized charge, spin and heat pumping in the adiabatic limit}

We start by considering the adiabatic pumping in a system where we have two regions 
\begin{equation}
m_0(x)=\begin{cases} 
0, & x < 0 \\
m_0, & x>0
\end{cases}
\end{equation}
and
\begin{equation}
\mu(x)=\begin{cases} 
\mu_{_{QSHI}}, & x < 0 \\
\mu_{_{FI}}, & x>0 
\end{cases}
\,.
\end{equation}
The exact conditions for the adiabatic limit to be valid are discussed below, but in this section we assume conservatively that $k_B T, |\mu_{_{FI}}|, \hbar |d\phi/dt|, \hbar |d\theta/dt| \ll m_0 \sin \theta$.

In the first lead, i.e.~quantum spin Hall insulator (QSHI) region $x<0$, the instantaneous scattering states of Eq.~(\ref{Normal-supp}) at energy $E$ can be written as
\begin{equation}
\psi_{_{QSHI}}(x) =  A_i\left ( 
\begin{array}{c} 
1 \\ 0  
\end{array}\right)e^{ik_1 x} +
A_o \left ( 
\begin{array}{c} 
0 \\ 1
\end{array}\right)e^{-ik_2 x}\,,
\end{equation}
where $k_1=\frac{E + \mu_{_{QSHI}}}{\hbar v_F}$ and $k_2=\frac{E + \mu_{_{QSHI}}}{\hbar v_F}$, and $A_i$ and $A_o$ are the amplitudes of the incoming and outgoing electrons.

In the ferromagnetic insulator (FI) region, $x>0$, 
the solution of Hamiltonian~(\ref{Normal-supp}) at energy $E$ satisfying $|E+\mu_{_{FI}}| < m_0 \sin\theta$ can be written as
\begin{equation}
\psi_{_{FI}}(x) =  C \left ( 
\begin{array}{c} 
E+\mu_{_{FI}}+i \sqrt{m_0^2 \sin^2\theta-(E+\mu_{_{FI}})^2} \\ m_0 \sin \theta e^{i\phi} 
\end{array}\right)e^{i\kappa x}\,,
\end{equation}
where $C$ is a constant and 
\begin{equation}
\kappa=-\frac{m_0 \cos \theta -\hbar \omega/2}{\hbar v_F}+i \sqrt{\frac{m_0^2 \sin^2\theta -(E+\mu_{_{FI}})^2}{\hbar^2 v_F^2}}\,.
\end{equation}
By matching the solutions at the boundary, we obtain the reflection coefficient
\begin{equation}
r_{ee}^{\downarrow\uparrow}(E)=\frac{A_o}{A_i}=\frac{m_0 \sin\theta e^{i \phi}}{E+\mu_{_{FI}}+i \sqrt{m_0^2 \sin^2\theta-(E+\mu_{_{FI}})^2}}\,. \label{scatt-simple}
\end{equation}
The pumped charge over one cycle and in the adiabatic limit can be calculated  from the expression 
\begin{equation}
Q_e =-\frac{e}{2\pi} \int\!dE \Bigl(\frac{\partial f}{\partial E}\Bigr)\!
\int_0^{\mathcal{T}} \!dt\, \mathrm{Im}\Bigl\{\mathrm{Tr}\Bigr[ \mathcal{S}^{\dagger}  \frac{\partial \mathcal{S}}{\partial t} \Bigr]\Bigr\}= \frac{ie}{2\pi} \int\!dE \Bigl(\frac{\partial f}{\partial E}\Bigr)\!
\int_0^{\mathcal{T}} \!dt\, [r_{ee}^{\downarrow\uparrow}(E)]^* \frac{\partial r_{ee}^{\downarrow\uparrow}(E)}{\partial t}\,,
\end{equation}
where $f(E)$ is the Fermi distribution function and 
\begin{equation}
    \mathcal{S}(E) = 
    \begin{pmatrix}
        0                       & 0 \\
    r_{ee}^{\downarrow\uparrow}(E) & 0
    \end{pmatrix}
\end{equation}
is the instantaneous scattering matrix pertaining to a normal metal-FI junction.

It is easy to see from Eq.~(\ref{scatt-simple}) that for $|E+\mu_{_{FI}}| < m_0 \sin\theta$ the magnitude of the reflection coefficient satisfies  $|r_{ee}^{\downarrow\uparrow}(E)|=1$ for all $0<\theta<\pi$ and all $\phi$. Notice that magnetization trajectory must satisfy $\theta(t) \ne 0,\pi$ for all $t$ because otherwise adiabaticity condition cannot be satisfied. Thus, we can write the reflection coefficient as 
\begin{equation}
r_{ee}^{\downarrow\uparrow}(E)=e^{i\phi} e^{i F(\theta, E)}
\end{equation}
to obtain
\begin{equation}
Q_e = -\frac{e}{2\pi} \int\!dE \Bigl(\frac{\partial f}{\partial E}\Bigr)\!
\int_0^{\mathcal{T}} \!dt\, \bigg[\frac{d\phi}{dt}+\frac{dF(\theta, E)}{dt}\bigg]=\frac{e}{2\pi} 
\int_0^{\mathcal{T}} \!dt\, \frac{d\phi}{dt}=\frac{e}{2\pi}[\phi(T)-\phi(0)]=eN,
\end{equation}
where $N\in \mathbb{Z}$ is the winding associated with the in-plane magnetization components over the cycle. The pumped charge is always an integer multiple of $e$ and independent of the trajectory of $\theta(t)$.

The spin current is trivially connected to the charge current so that the spin pumped over the cycle is obtained just by replacing $e \to - \hbar/2$, and therefore it is also quantized and independent of $\theta(t)$ as long as the adiabaticity conditions are satisfied.

We now turn our attention to the pumped heat during the cycle. It is described by 
\begin{eqnarray}
Q_E &=& -\frac{\hbar}{4\pi} \int\!dE \Bigl(\frac{\partial f}{\partial E}\Bigr)\!
\int_0^{\mathcal{T}} \!dt\, \mathrm{Tr}\Bigr[\frac{\partial \mathcal{S}}{\partial t}\frac{\partial \mathcal{S}^{\dagger}}{\partial t}\Bigr]=-\frac{\hbar}{4\pi} \int\!dE \Bigl(\frac{\partial f}{\partial E}\Bigr)\!
\int_0^{\mathcal{T}} \!dt\, \frac{\partial [r_{ee}^{\downarrow\uparrow}(E)]^*}{\partial t}\frac{\partial r_{ee}^{\downarrow\uparrow}(E)}{\partial t} \nonumber \\ 
&=&-\frac{\hbar}{4\pi} \int\!dE \Bigl(\frac{\partial f}{\partial E}\Bigr)\!
\int_0^{\mathcal{T}} \!dt\, \bigg[\frac{d\phi}{dt}+\frac{dF(\theta, E)}{dt}\bigg]^2.
\end{eqnarray}
From this expression it is clear that $Q_E$ depends on the trajectory. For circular precession $\phi(t)=\omega t$ and time-independent $\theta(t)=\theta$, we obtain
\begin{eqnarray}
Q_E = \frac{\hbar \omega}{2}.
\end{eqnarray}
Thus the value of the pumped heat is still robust in the sense that it does not depend on the precession angle $\theta$.

\subsection{Connection to the Chern number in the case of Hamiltonian formalism}

Next we analyze the low-energy Hamiltonian pertaining to the edges in the case of an infinite system.  For that, we first write the  Hamiltonian describing the edge modes in the presence of an arbitrary  magnetization as follows:
\begin{align}
    H(p, t)&=v_F[p+m_z(t)/v_F]\sigma_z+m_x(t)\sigma_x+m_y(t)\sigma_y\,,
    \label{Ham}
\end{align}
In this case, the charge current density can be written as:
\begin{align}
    j_c(p, t)= ev_F\langle\sigma_z(p, t)\rangle/L\,,  
    \label{current}
\end{align}
with $L$ being the system size and the expectation value $\langle\dots\rangle$ being taken over the stationary (but not equilibrium) state of the system. Next we evaluate the expectation value $\langle\sigma_z(p, t)\rangle$ using a time-dependent adiabatic perturbation theory justified by the presence of the gap $\Delta_F(t)=\sqrt{m_x^2(t)+m_y^2(t)}$ in the spectrum.  

We perform a time-dependent unitary transformation $\mathcal{U}(p,t)$ that diagonalizes the instantaneous Hamiltonian $H(p, t)$ at the expense of introducing non-diagonal gauge fields terms:
\begin{align}
    \widetilde{H}(p,t)&=\mathcal{U}^\dagger(t) H_{edge}(p,t)\mathcal{U}(t)-i\hbar \mathcal{U}^\dagger(p,t)\partial_t\mathcal{U}(p,t)\equiv\epsilon(p,t)\sigma_z+\hbar \dot{\bs m}(t)\cdot{\bs A}_{\bs m}(p,t)\,,\\    
    \epsilon(p,t)&=\sqrt{[v_Fp+m_z(t)]^2+m_x^2(t)+m_y^2(t)}\,,
\end{align}
where ${\bs A}_{\bs m}(p,t)=-i  \mathcal{U}^\dagger(p, t)\partial_{{\bs m}}\mathcal{U}(p, t)$ is the corresponding Berry connection vector. Note that in the new basis the spin operator is $\tilde{\sigma}_z(p,t)=\mathcal{U}^\dagger(p, t)\sigma_z(p,t)\mathcal{U}(p, t)$. Using the  relationship $\sigma_\alpha(p, t )=\partial_{m_\alpha} H(p, t)$, we obtain
\begin{eqnarray}
\tilde{\sigma}_\alpha(p,t)&=&\mathcal{U}^\dagger(t) \big( \partial_{m_\alpha} H_{edge} \big) \mathcal{U}(t)=\partial_{m_\alpha} \big( \mathcal{U}^\dagger(t)  H_{edge} \mathcal{U}(t) \big)- \big(\partial_{m_\alpha} \mathcal{U}^\dagger(t) \big)  H_{edge} \mathcal{U}(t)-\mathcal{U}^\dagger(t)  H_{edge} \big( \partial_{m_\alpha}   \mathcal{U}(t) \big) \nonumber \\
&=&\partial_{m_\alpha} \epsilon(p,t) \sigma_z +  i \epsilon(p,t) [A_{m_\alpha}(p,t),  \sigma_z]. 
\end{eqnarray}

Next we treat  the gauge field term in (time-dependent) perturbation theory, by keeping only the terms that are leading order in the velocities $\dot{m}_\alpha$. That can be implemented using another time-dependent unitary transformation, $\mathcal{U}'(p,t)=e^{-S(p, t)}=1-S(p, t)+\dots$, with $S^\dagger(p, t)=-S(p, t)$, chosen such  that
\begin{align}
S(p, t)=
\frac{\hbar}{4\epsilon(p,t)} [\sigma_z, \dot{\bs m}(t)\cdot{\bs A}_{\bs m}(p,t)],
\end{align}
which in turn affects the spin operator  $\tilde{\sigma}_z(p, t)\rightarrow\tilde{\sigma}_z(p, t)+[S(p, t),\tilde{\sigma}_z(p, t)]$, in leading order in velocities. 
We can now readily evaluate the expectation value of $\sigma_z(p,t)$ for the eigenstates of the  Hamiltonian:
\begin{align}
    \langle\sigma_z(p,t)\rangle_\sigma\approx\langle\sigma|\tilde{\sigma}_z(p,t)|\sigma\rangle+\langle\sigma|[S(p, t),\tilde{\sigma}_z(p, t)]|\sigma\rangle=\langle\sigma_z(p, t)\rangle_{\sigma,i}+\sum_{\alpha=x,y}\hbar \dot{m}_\alpha(t)\mathcal{F}^\sigma_{\alpha z}(p, t)\,,
\end{align}
$\langle\sigma_z(p, t)\rangle_{\sigma,i}$ being the instantaneous spin expectation value in eigenstate $|\sigma\rangle$ of Hamiltonian (\ref{Ham}), and  $\mathcal{F}^\sigma_{\alpha\beta}(p, t)=i\langle\sigma|[A_{m_\alpha}(p, t),A_{m_\beta}(p, t)]|\sigma\rangle$ being the Berry curvature components pertaining to the state $|\sigma\rangle$ and momentum $p$, with $\alpha,\beta=x,y,z$. To evaluate the total spin of the electrons, we need to account for all the occupied  states up to the Fermi energy, so that we get:
\begin{align}
\langle\sigma_z(t)\rangle/L=\frac{1}{2\pi \hbar}\int_{-\infty}^\infty dp\langle\sigma_z(p, t)\rangle=-\dot{m}_x(t)\mathcal{F}^\downarrow_{zx}(t)+ \dot{m}_y(t)\mathcal{F}^\downarrow_{yz}(t)\,,
\end{align}
where $\mathcal{F}^\downarrow_{\alpha\beta}(t)=\frac{1}{2\pi}\int_{-\infty}^\infty dp\, \mathcal{F}^\downarrow_{\alpha\beta}(p, t)$. Here, we have assumed that the instantaneous contribution vanishes upon integration. We point out that the appearance of the instantaneous expectation value for $\sigma_z$ depends on the regularization of the integration limits, and indeed it can be non-zero because we have magnetization along the $z$-direction. Nevertheless, we are here ultimately interested about the charge current, which is related to expectation value of $\sigma_z$ within the low-energy theory, but the instantaneous expectation value of the charge current operator must always vanish for fully occupied bands also in the case of a full lattice model. Thus, it is well-justified to neglect the instantaneous expectation value of $\sigma_z$ in the context of our theory.

The general expression for the Berry curvature components for a Hamiltonian $H={\bs b}(p, t) \cdot {\bs \sigma}$ reads 
\begin{align}
    \mathcal{F}^\downarrow_{\alpha\beta}(p, t)&=i\langle\downarrow|[A_{m_\alpha}(p,t),A_{m_\beta}(p,t)]|\downarrow\rangle=\epsilon_{\alpha\beta\gamma}\frac{b_{\gamma}(p, t)}{2\epsilon^{3}(p, t)}\,,
\end{align}
which leads to the following integrals:
\begin{align}
    \mathcal{F}_{zx}^\downarrow(t)&=\frac{m_y(t)}{4\pi}\int_{-\infty}^\infty dp\frac{1}{\epsilon^3(p, t)}=\frac{m_y(t)}{2\pi v_F [m_x^2(t)+m_y^2(t)]}\,,\\
    \mathcal{F}_{yz}^\downarrow(t)&=\frac{m_x(t)}{4\pi}\int_{-\infty}^\infty dp\frac{1}{\epsilon^3(p, t)}=\frac{m_x(t)}{2\pi v_F [m_x^2(t)+m_y^2(t)]}\,.
\end{align}
Inserting these expressions in the formula for the spin expectation value and assuming that $m_x(t)=m_{xy}(t) \cos\phi(t)$ and $m_y(t)=m_{xy}(t) \sin\phi(t)$ we obtain:
\begin{align}
    \langle\sigma_z(t)\rangle/L&=  \frac{-\dot{m}_x(t)m_y(t)+\dot{m}_y(t)m_x(t)}{2\pi v_F[m_x^2(t)+m_y^2(t)]}\equiv \frac{1}{2\pi v_F}\dot{\phi}(t)\,.
\end{align}
We are now in position to evaluate the expectation value for the charge current and, moreover, the charge transferred between the sources by this pump (or the charge that passes through an arbitrary point in the system during one period $\mathcal{T}$):
\begin{align}
    Q_e=\int_{0}^{\mathcal{T}}dtj_c(t)=\frac{e}{2\pi}[\phi(T)-\phi(0)]\equiv eN\,,
\end{align}
where $N\in \mathbb{Z}$ is the winding number associated with the in-plane magnetization components. For the circular driving presented in the paper, we obtain $N=1$, which proves our scattering matrix results and it is consistent with the numerical calculations.

In the calculation of the pumped charge we ended up integrating the Berry curvature over $p$ and $\phi$. Thus, if these variables are periodic (or can be compactified), the pumped charge equals the Chern number, and must therefore always be an integer. To make this connection more explicit we consider an arbitrary lattice Hamiltonian of the form
\begin{equation}
    \mathcal{H}(k, t) = \mathcal{H}_0(k) +{\bs m}(t)\cdot{\bs \sigma}, 
\end{equation}
where ${\bs m}(t)=m_0[\sin{\theta(t)}\cos\phi(t),\sin{\theta(t)} \sin{\phi(t)},\cos{\theta(t)}]$, and $\mathcal{H}_0(k)$ satisfies periodicity $\mathcal{H}_0(k+2\pi)=\mathcal{H}_0(k)$. We assume that Hamiltonian $\mathcal{H}(k, t)$ is gapped so that we can consider adiabatic time-evolution.

The time-dependent current expectation value is 
\begin{equation}
j_c(t)=\frac{1}{2\pi} \int_0^{2\pi} dk~\langle j_c(k,t) \rangle_{s},
\end{equation}
where the current density operator is defined as 
\begin{equation}
j_c(k,t)=\frac{e}{\hbar} \partial_{k}\mathcal{H}(k, t)
\end{equation}
and the average being taken over the time-dependent state of the system. We perform a  unitary transformation $\mathcal{U}(k, t)$ that diagonalizes the instantaneous Hamiltonian $\mathcal{H}(k, t)$ at the expense of introducing non-diagonal gauge fields terms
\begin{align}
    \widetilde{\mathcal{H}}(k, t)&=\mathcal{U}^\dagger(t) \mathcal{H}(k, t)\mathcal{U}(t)-i\hbar \mathcal{U}^\dagger(k, t)\partial_t\mathcal{U}(k, t)\equiv\sum_n\epsilon_n(k,t) |n k\rangle\langle n k|+\hbar \dot{\bs m}(t)\cdot{\bs A}_{\bs m}(k, t)\,,    
\end{align}
where $\epsilon_{n}(k, t)$ and $|nk \rangle$ label the eigenenergies and eigenstates with band index $n$ and momentum $k$, and ${\bs A}_{\bs m}(k,t)=-i\mathcal{U}^\dagger(k, t)\partial_{{\bs m}}\mathcal{U}(k, t)$ is the Berry connection vector associated with the magnetization dynamics. Note that after the transformation, the charge current operator $j_c(k, t)\rightarrow \tilde{j}_c(k, t)=\mathcal{U}^\dagger(k,t) j_c(k, t)\mathcal{U}(k,t)$. Next we treat the gauge field term in (time-dependent) perturbation theory, keeping only the terms that are leading order in the velocities $\dot{m}_\alpha$. That can be implemented using another time-dependent unitary transformation, $\mathcal{U}'( k,t)=e^{-S(k, t)}=1-S(k, t)+\dots$, with $S^\dagger(k, t)=-S(k, t)$, chosen such  that
\begin{equation}
S(k, t)=\hbar \sum_{n \ne n'} \frac{\dot{\bs m}(t)\cdot\langle n k|{\bs A}_{\bs m}(k, t)|n' k\rangle}{\epsilon_n(k, t)-\epsilon_{n'}(k, t)} |nk \rangle \langle n'k |=i\hbar \sum_{n \ne n'} \frac{\dot{\bs m}(t)\cdot\langle nk| \mathcal{U}^\dagger(k, t) \partial_{\bs m}\mathcal{H}(k, t) \mathcal{U}(k, t) |n' k\rangle}{[\epsilon_n(k, t)-\epsilon_{n'}(k, t)]^2}|nk \rangle \langle n'k |,
\end{equation}
which affects the current operator as $\tilde{j}_c(k, t)\rightarrow \tilde{j}_c(k, t)+[S(k, t), \tilde{j}_c(k, t)]$ in leading order in velocities. Note that the diagonal terms of the gauge fields only renormalize the energy differences, and this effect can be neglected when keeping only the terms which are leading order in velocities.  Utilising the above expressions, we can now evaluate the current expectation value in leading order in velocities
\begin{align}
    j_c(t)&=\frac{e}{h}\sum_{n\in occ} \int_0^{2\pi} dk~\widetilde{\langle n k,t|}\partial_k\mathcal{H}(k, t) \widetilde{|n k,t\rangle}\nonumber\\
    &+ \frac{e}{\pi} {\rm Im}\sum_{n\in occ, n'\in empty} \int_0^{2\pi} dk \displaystyle{\frac{\dot{\bs m}(t)\cdot\widetilde{\langle n k,t|}\partial_k\mathcal{H}(k, t)\widetilde{|n' k,t\rangle}\widetilde{\langle n' k,t|}\partial_{\bs m}\mathcal{H}(k, t)\widetilde{|n k,t\rangle}}{[\epsilon_n(k, t)-\epsilon_{n'}(k, t)]^2}},
\end{align}
where $\widetilde{|n k, t\rangle}=\mathcal{U}(k, t)|n k\rangle$ are the instantaneous eigenstates in the original (lab) frame. The first term always vanishes for fully occupied bands. Thus, the charge pumped over one cycle is given by 
\begin{equation}
    Q_e=\int_0^{\mathcal{T}}dtj_c(t)=\frac{e}{\pi} {\rm Im}\sum_{n\in occ, n'\in empty} \int_0^{\mathcal{T}}dt \int_0^{2\pi} dk \displaystyle{\frac{\dot{\bs m}(t)\cdot\widetilde{\langle n k,t|}\partial_k\mathcal{H}(k, t)\widetilde{|n' k,t\rangle}\widetilde{\langle n' k,t|}\partial_{\bs m}\mathcal{H}(k, t)\widetilde{|n k,t\rangle}}{[\epsilon_n(k, t)-\epsilon_{n'}(k, t)]^2}}.
\end{equation}
Assuming that $\theta(t)=\theta$ is time-independent and $\phi(t)$ goes from $0$ to $2\pi$ during one cycle, we obtain
\begin{equation}
    Q_e=e \frac{1}{2\pi} \sum_{n\in occ, n'\in empty} \int_0^{2\pi}d\phi \int_0^{2\pi} dk \ {\rm Im} \bigg\{ \frac{2 \widetilde{\langle n k,\theta, \phi|}\partial_k\mathcal{H}(k, \theta, \phi)\widetilde{|n' k, \theta, \phi\rangle}\widetilde{\langle n' k,\theta, \phi|}\partial_\phi\mathcal{H}(k, \theta, \phi)\widetilde{|n k, \theta, \phi\rangle}}{[\epsilon_n(k, \theta, \phi)-\epsilon_{n'}(k, \theta,  \phi)]^2} \bigg\}.
\end{equation}
Therefore, $Q_e=e C$, where $C$ is the Chern number in the two-dimensional ($k$, $\phi$)-space. The Chern number $C$ must always be an integer and it can only change if there is a gap closing in the Hamiltonian. Thus the pumped charge is independent of $\theta$ in a region of $\theta$ where there is no gap closings in the Hamiltonian. Moreover, it also does not depend on the time-dependent trajectory $\theta(t)$  as long as gap closing points are not crossed during the deformations of $\theta(t)$. In our model, the gap closings take place at $\theta=0$  and $\theta =\pi$. Thus the pumped charge does not depend on $\theta(t)$ as long as $0<\theta(t)<\pi$ and the pumped charge is determined by the winding of $\phi(t)$ during the cycle. If the winding of $\phi(t)$ changes the trajectory must cross the point $\theta=0$ or $\theta=\pi$ during the deformation.

\subsection{Frequency-dependence of the pumped current}

The discussion above gives an impression that the pumped currents do not depend on $\theta$. However, this is only true in the adiabatic limit. To study the frequency-dependence of the pumped currents, we consider Hamiltonian (\ref{Normal-supp}), where
${\bs m}(x,t)=m_0(x)[\sin{\theta}\cos(\omega t),\sin{\theta} \sin(\omega t),\cos{\theta}]$, and we separate the system in three regions 
\begin{equation}
m_0(x)=\begin{cases} 
0, & x < 0 \\
m_0, & 0<x<L \\
0, & x>L
\end{cases}
\end{equation}
and
\begin{equation}
\mu(x)=\begin{cases} 
\mu_{_{QSHI}}, & x < 0 \\
\mu_{_{FI}}, & 0<x<L \\
\mu_{_{QSHI}}, & x > L
\end{cases}
\,.
\end{equation}

We perform a unitary transformation $U=e^{-i\omega t \sigma_z/2}$ to the rotating frame (in spin-space), which brings the Hamiltonian in the time-independent form
\begin{equation}
H_{\rm eff}=U^\dag H(t) U-i\hbar U^\dag \frac{\partial U}{\partial t}=v_Fp\,\sigma_z-\mu(x)+m_0(x) \cos \theta \sigma_z+m_0(x) \sin \theta \sigma_x  -\frac{\hbar \omega}{2} \sigma_z\,, \label{effective-Hamiltonian-normal}
\end{equation}
but there now exists a fictitious spin bias term $-\frac{\hbar \omega}{2} \sigma_z$.  Due to the non-equilibrium nature of the problem this term acts not only on the spectrum but it also shifts the distribution functions of the spin up and spin down electrons in leads. Otherwise, the transport in this system system can be calculated in similar way as in the case of static Hamiltonians. Therefore, we start by computing the reflection coefficient for the Hamiltonian (\ref{effective-Hamiltonian-normal}).

In the first lead, i.e.~QSHI region $x<0$, the solution of Eq.~(\ref{effective-Hamiltonian-normal}) at energy $E$ can be written as
\begin{equation}
\psi^1_{_{QSHI}}(x) =  A_i\left ( 
\begin{array}{c} 
1 \\ 0  
\end{array}\right)e^{ik_1 x} +
A_o \left ( 
\begin{array}{c} 
0 \\ 1
\end{array}\right)e^{-ik_2 x}\,,
\label{wfQSHI}
\end{equation}
where $k_1=\frac{E + \mu_{_{QSHI}}+\hbar\omega/2}{\hbar v_F}$ and $k_2=\frac{E + \mu_{_{QSHI}}-\hbar\omega/2}{\hbar v_F}$, and the coefficients $A_i$ and $A_o$ are the amplitudes of the incoming and outgoing electrons.

Similarly in the second lead, QSHI region $x>L$, the solution of Eq.~(\ref{effective-Hamiltonian-normal}) at energy $E$ can be written as
\begin{equation}
\psi^2_{_{QSHI}}(x) =  B_o\left ( 
\begin{array}{c} 
1 \\ 0  
\end{array}\right)e^{ik_1 (x-L)} +
B_i \left ( 
\begin{array}{c} 
0 \\ 1
\end{array}\right)e^{-ik_2 (x-L)}\,,
\label{wfQSHI}
\end{equation}
where the coefficients $B_i$ and $B_o$ are the amplitudes of the incoming and outgoing electrons.

In the FI region, $0<x<L$, we have to consider the cases (i) $(E+\mu_{_{FI}})^2 < m_0^2 \sin^2\theta$ and (ii) $(E+\mu_{_{FI}})^2 > m_0^2 \sin^2\theta$ separately.

(i) For $(E+\mu_{_{FI}})^2 < m_0^2 \sin^2\theta$ 
the solution of Eq.~(\ref{effective-Hamiltonian-normal}) at energy $E$ can be written as
\begin{equation}
\psi_{_{FI}}(x) =  C_1 \left ( 
\begin{array}{c} 
E+\mu_{_{FI}}+i \sqrt{m_0^2 \sin^2\theta-(E+\mu_{_{FI}})^2} \\ m_0 \sin \theta  
\end{array}\right)e^{i\kappa_1 x}+C_2 \left ( 
\begin{array}{c} 
E+\mu_{_{FI}}-i \sqrt{m_0^2 \sin^2\theta-(E+\mu_{_{FI}})^2} \\ m_0 \sin \theta  
\end{array}\right)e^{i\kappa_2 x}\,,
\label{wfFI1}
\end{equation}
where $C_1$ and $C_2$ are constants, 
\begin{equation}
\kappa_1=-\frac{m_0 \cos \theta -\hbar \omega/2}{\hbar v_F}+i \sqrt{\frac{m_0^2 \sin^2\theta -(E+\mu_{_{FI}})^2}{\hbar^2 v_F^2}}
\end{equation}
and 
\begin{equation}
\kappa_2=-\frac{m_0 \cos \theta -\hbar \omega/2}{\hbar v_F}-i \sqrt{\frac{m_0^2 \sin^2\theta -(E+\mu_{_{FI}})^2}{\hbar^2 v_F^2}}.
\end{equation}
To solve the reflection coefficient we assume that the electrons are incoming from lead 1 so that $B_i=0$. Then we obtain
\begin{equation}
r_{ee}^{\downarrow\uparrow}(E)=\frac{A_o}{A_i}=\frac{m_0 \sin\theta}{E+\mu_{_{FI}}+i \sqrt{m_0^2 \sin^2\theta-(E+\mu_{_{FI}})^2}}\frac{1-e^{i(\kappa_1-\kappa_2)L}}{1-\displaystyle{e^{i(\kappa_1-\kappa_2)L} \frac{m_0^2 \sin^2\theta}{[E+\mu_{_{FI}}+i \sqrt{m_0^2 \sin^2\theta-(E+\mu_{_{FI}})^2}]^2}}}\,.
\end{equation}
Notice that $|r_{ee}^{\downarrow\uparrow}(E)|=1$ up to exponentially small corrections $\propto \exp\bigg\{- 2\sqrt{\frac{m_0^2 \sin^2\theta -(E+\mu_{_{FI}})^2}{\hbar^2 v_F^2}} L\bigg\}$ associated with the electron tunneling across the FI region twice.

(ii) For $(E+\mu_{_{FI}})^2 > m_0^2 \sin^2\theta$ 
the solution of Eq.~(\ref{effective-Hamiltonian-normal}) at energy $E$ can be written as
\begin{equation}
\psi_{_{FI}}(x) =  D_1 \left ( 
\begin{array}{c} 
E+\mu_{_{FI}}+ \sqrt{(E+\mu_{_{FI}})^2-m_0^2 \sin^2\theta} \\ m_0 \sin \theta  
\end{array}\right)e^{iq_1 x}+D_2 \left ( 
\begin{array}{c} 
E+\mu_{_{FI}}- \sqrt{(E+\mu_{_{FI}})^2-m_0^2 \sin^2\theta} \\ m_0 \sin \theta  
\end{array}\right)e^{iq_2 x}\,,
\label{wfFI2}
\end{equation}
where $D_1$ and $D_2$ are constants,  
\begin{equation}
q_1=-\frac{m_0 \cos \theta -\hbar \omega/2}{\hbar v_F}+ \sqrt{\frac{(E+\mu_{_{FI}})^2-m_0^2 \sin^2\theta}{\hbar^2 v_F^2}}
\end{equation}
and
\begin{equation}
q_2=-\frac{m_0 \cos \theta -\hbar \omega/2}{\hbar v_F}- \sqrt{\frac{(E+\mu_{_{FI}})^2-m_0^2 \sin^2\theta}{\hbar^2 v_F^2}}\,.
\end{equation}
To solve the reflection coefficient we assume that the electrons are incoming from lead 1 so that $B_i=0$. Then we obtain
\begin{equation}
r_{ee}^{\downarrow\uparrow}(E)=\frac{A_o}{A_i}=\frac{m_0 \sin \theta}{E+\mu_{_{FI}}+\sqrt{(E+\mu_{_{FI}})^2-m_0^2 \sin^2\theta}}\frac{1-e^{i(q_1-q_2)L}}{1-e^{i(q_1-q_2)L} \frac{m_0^2 \sin^2\theta}{[E+\mu_{_{FI}}+\sqrt{(E+\mu_{_{FI}})^2-m_0^2 \sin^2\theta}]^2}}\,.
\end{equation}

In the following, we are only interested about absolute value of the reflection coefficient
\begin{equation}
|r_{ee}^{\downarrow\uparrow}(E)|^2=\begin{cases} \frac{\sinh^2\bigg[\sqrt{\frac{m_0^2 \sin^2\theta-(E+\mu_{_{FI}})^2}{\hbar^2 v_F^2}}L \bigg]}{\sinh^2\bigg[\sqrt{\frac{m_0^2 \sin^2\theta-(E+\mu_{_{FI}})^2}{\hbar^2 v_F^2}}L \bigg]+\frac{m_0^2 \sin^2\theta-(E+\mu_{_{FI}})^2}{m_0^2 \sin^2\theta}}
=\frac{\sinh^2\big[\frac{L}{\xi_F(E+\mu_{_{FI}}, \theta)} \big]}{\sinh^2\big[\frac{L}{\xi_F(E+\mu_{_{FI}}, \theta)} \big]+\frac{\xi_F^2(0, \theta)}{\xi_F^2(E+\mu_{_{FI}}, \theta)}}, & |E+\mu_{_{FI}}|< m_0 \sin\theta \\\\
\frac{\sin^2\bigg[\sqrt{\frac{(E+\mu_{_{FI}})^2-m_0^2 \sin^2\theta}{\hbar^2 v_F^2}}L \bigg]}{\sin^2\bigg[\sqrt{\frac{(E+\mu_{_{FI}})^2-m_0^2 \sin^2\theta}{\hbar^2 v_F^2}}L \bigg] +\frac{(E+\mu_{_{FI}})^2-m_0^2 \sin^2\theta}{m_0^2 \sin^2\theta}}=\frac{\sin^2\big[\frac{L}{\xi_F(E+\mu_{_{FI}}, \theta)} \big]}{\sin^2\big[\frac{L}{\xi_F(E+\mu_{_{FI}}, \theta)} \big]+\frac{\xi_F^2(0, \theta)}{\xi_F^2(E+\mu_{_{FI}}, \theta)}},& |E+\mu_{_{FI}}| > m_0 \sin\theta, \end{cases}
\end{equation}
where $\xi_F(E, \theta)=\hbar v_F/\sqrt{|m_0^2 \sin^2\theta-E^2|}$.

We are now ready to proceed to the calculation of pumped currents. For simplicity in this section we  assume that the drain voltage $V_d=0$, and we concentrate on charge current although similar results can be derived also for spin and heat currents. By including the scattering paths in the presence of fictitious bias  the charge current flowing into the normal lead can be written as 
\begin{equation}
\langle I_e \rangle = \frac{e}{h} \int dE \bigg[f\left(E-\frac{\hbar \omega}{2}\right)-f\left(E+\frac{\hbar \omega}{2}\right)\bigg]
|r_{ee}^{\downarrow\uparrow}(E)|^2\,.
\end{equation}

Assuming that $|\mu_{_{FI}}|, \hbar \omega, k_B T \ll m_0 \sin\theta$ and $L \gg  \xi_{F} (0,\theta)$, we obtain the quantized charge current
\begin{equation}
\langle I_e \rangle = \frac{e \omega}{2\pi}\,.
\end{equation}

\begin{figure}
    \centering
     \includegraphics[width=0.45\linewidth]{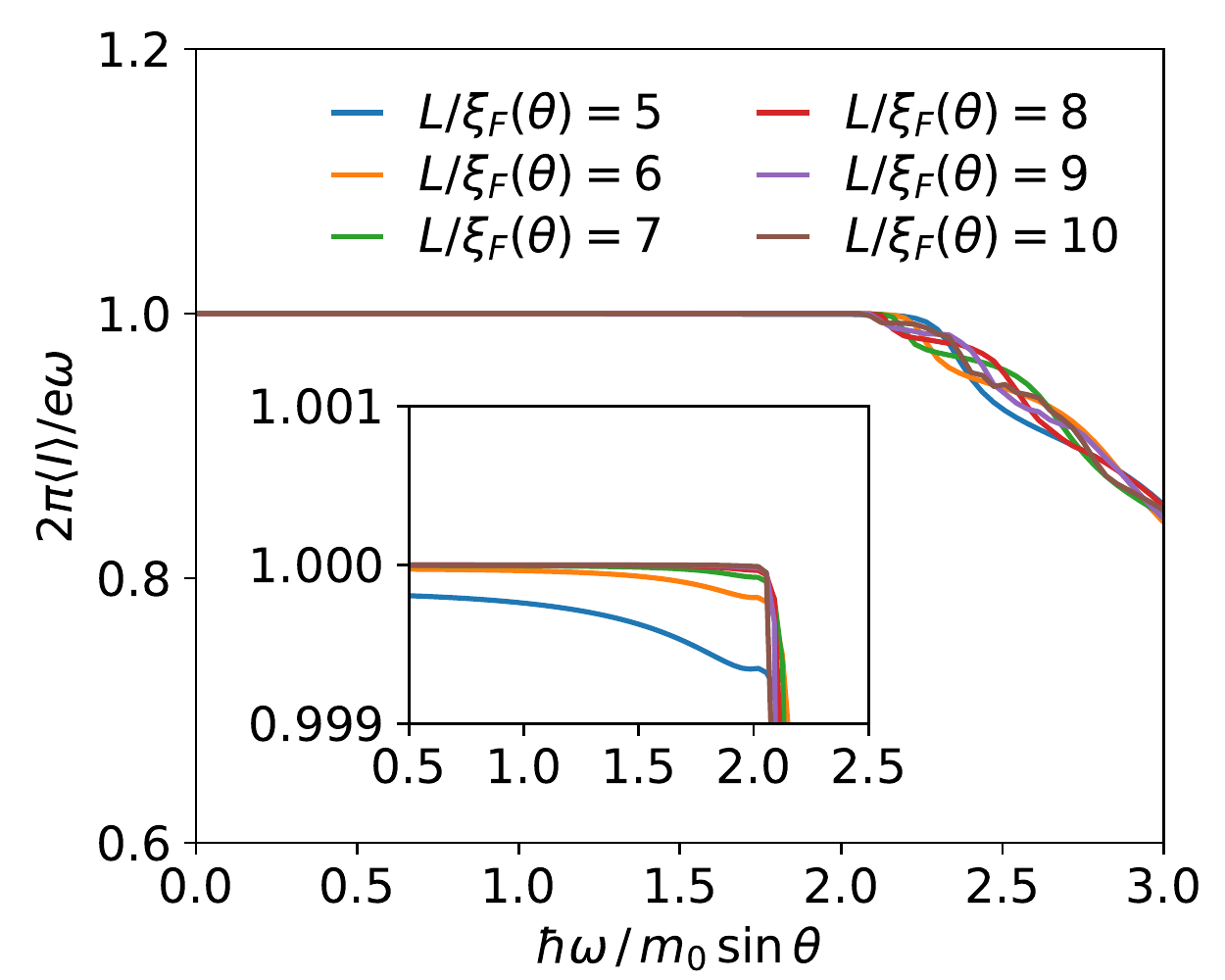}
    \caption{Pumped charge $Q_e=2 \pi \langle I_e \rangle/e\omega$ as a function of $\hbar\omega/m_0\sin{\theta}$ for $\mu_{_{FI}}=k_B T=0$ and various values of $L/\xi(\theta)$, where $\xi(\theta)=\hbar v_F/|m_0\sin{\theta}|$ is the FI coherence length. The pumped charge is quantized to $Q_e=1$ for $\hbar\omega/m_0\sin{\theta}<2$.} 
    \label{QvsFreq_1d&lat}
\end{figure}

In the limit $k_B T=0$, the expression simplifies to 
\begin{equation}
\langle I_e \rangle = \frac{e}{h} \int^{\hbar \omega/2}_{-\hbar \omega/2} dE 
|r_{ee}^{\downarrow\uparrow}(E)|^2\,.
\end{equation}
From this expression we see that in the limit $L \gg  \xi_F(0, \theta)$, the quantization breaks down only if 
\begin{equation}
\hbar\omega > 2(m_0 \sin \theta-|\mu_{_{FI}}|)\,.
\label{quantibreak}
\end{equation}
From this expression it is easy to see that once $\theta$ approaches $0, \pi$ the quantization occurs in a narrower range of frequencies. The frequency dependence in the case of $\mu_{_{FI}}=0$ is shown in Fig.~\ref{QvsFreq_1d&lat} for various values of $L/\xi_F(0,\theta)$.

\section{\label{sec:level1}Analytical calculation of the instantaneous scattering matrix in the presence of superconductivity}

\subsection{\label{app:subsec}The instantaneous eigenstates}

In this section we calculate the instaneous scattering states of the BdG hamiltonian
\begin{align}
    \mathcal{H}(t)=[v_Fp\,\sigma_z-\mu(x)]\tau_z+{\bs m}(x,t)\cdot{\bs \sigma}+\Delta(x)\tau_x \label{BdG-supp}
\end{align}
considered in the main text. This Hamiltonian obeys a particle-hole symmetry
\begin{equation}
\tau_y \sigma_y H^T(-p) \tau_y \sigma_y=-H(p).
\end{equation}
For this purpose, we separate the system into three regions depending on whether the induced magnetism/superconductivity is present. 

In region I, ${\bs m}=\Delta=0$, so that the solutions of Eq.~(\ref{BdG-supp}) at energy $E$ are
\begin{equation}
\psi_I(x) =  A_e^i\left ( 
\begin{array}{c} 
1 \\ 0 \\ 0 \\ 0 
\end{array}\right)e^{ik_1 x} +
A_e^o \left ( 
\begin{array}{c} 
0 \\ 1 \\ 0 \\ 0 
\end{array}\right)e^{-ik_1 x} +
A_h^o \left ( 
\begin{array}{c}
0 \\ 0 \\ 1 \\ 0 
\end{array}\right)e^{ik_2 x} +
A_h^i \left ( 
\begin{array}{c}
0 \\ 0 \\ 0 \\ 1 
\end{array}\right)e^{-ik_2 x}\,,
\label{wfncinI}
\end{equation}

\noindent where $k_1=\frac{E + \mu_{_{QSHI}}}{\hbar v_F}$, $k_2=\frac{\mu_{_{QSHI}} - E}{\hbar v_F}$, and $\mu_{_{QSHI}}$ is the chemical potential in region I.
The coefficients $A_e^o$ and $A_h^o$ determine the amplitude of outgoing electrons and holes with spin down, respectively. On the other hand, the coefficients $A_e^i$ and $A_h^i$ describe the amplitudes of incident electrons and holes with spin up, respectively. 

In region II, ${\bs m} = m_0 [\sin \theta \cos \phi, \sin\theta \sin \phi, \cos\theta]$ and $\Delta=0$, so that a proximity-induced magnetism  opens an energy gap, and we assume that the Fermi level is always inside this gap $|E\pm \mu_{_{FI}}|< m_0\sin{\theta}$. Then, the solution of Eq.~(\ref{BdG-supp}) at energy $E$ reads
\begin{eqnarray}
\psi_{II}(x) &=& b 
\left ( 
\begin{array}{c} e^{i\gamma_m^+}e^{-i\phi} \\ 1 \\ 0 \\ 0 \end{array}\right)
e^{-ik_m x} e^{-q_e x} +
b' \left ( 
\begin{array}{c} e^{-i\gamma_m^+}e^{-i\phi} \\ 1 \\ 0 \\ 0 \end{array}\right)
e^{-ik_m x} e^{q_e x} \\ 
&+& d \left ( 
\begin{array}{c} 0 \\ 0 \\ e^{i\gamma_m^-}e^{-i\phi} \\ 1 \end{array}\right)
e^{ik_m x} e^{q_h x} +
d' \left ( 
\begin{array}{c} 0 \\ 0 \\ e^{-i\gamma_m^-}e^{-i\phi} \\ 1 \end{array}\right)
e^{ik_m x} e^{-q_h x}\,, \nonumber
\label{wfncinII}
\end{eqnarray}

\noindent where $e^{i\gamma_m^\pm}=(E\pm\mu_{_{FI}}+i \sqrt{m_0^2\sin^2{\theta} - (E\pm \mu_{_{FI}})^2})/(m_0\sin{\theta})$, $k_m = m_z/\hbar v_F$, $q_{e,h}=\sqrt{m_0^2\sin^2{\theta} - (E\pm \mu_{_{FI}})^2}/\hbar v_F$, and $m_z=m_0\cos{\theta}$. In the following, we denote $\mu_{_{FI}}=eV_g$ to indicate that the potential in this region can be controlled with a gate voltage.

In region III,  ${\bs m}=0$ and $\Delta=\Delta_0$, so that the proximity induced superconductivity opens an energy gap. Then, the exponentially decaying solution of Eq.~(\ref{BdG-supp}) at energy $E$ reads
\begin{equation}
\psi_{III}(x) = c \left ( 
\begin{array}{c} 
e^{i\delta} \\ 0 \\ 1 \\ 0
\end{array}
\right)e^{i \frac{\mu_{SC}}{\hbar v_F} x}e^{-\frac{\sqrt{\Delta_{0}^2-E^2}}{\hbar v_F}x} +
f \left ( 
\begin{array}{c}
0 \\ e^{-i\delta} \\ 0 \\ 1
\end{array}\right)e^{-i \frac{\mu_{SC}}{\hbar v_F} x} e^{-\frac{\sqrt{\Delta_{0}^2-E^2}}{\hbar v_F}x}\,, 
\label{wfncinIII}
\end{equation}
\noindent where $e^{i\delta}=(E+i\sqrt{\Delta_0^2-E^2)}/\Delta_0$.

\subsection{\label{app:subsec}The reflection coefficients}

The continuity of the wave function at the interfaces between the three different regions provides a set of linear equations that determines the unknown coefficients in Eqs.~(\ref{wfncinI})-(\ref{wfncinIII}).  After straightforward calculations for the case $A_h^i=0$, we obtain that the reflection coefficients $r_{ee}^{\downarrow\uparrow}(E) = A_e^o/A_e^i$ and $r_{he}^{\downarrow\uparrow}(E) = A_h^o/A_e^i$  are given by
\begin{eqnarray}
r_{ee}^{\downarrow\uparrow}(E) &=& e^{i\phi}e^{-2ik_1L}\frac{
e^{2i\delta}\bigl(\cos{\gamma_m^-} + i\sin{\gamma_m^-}\coth{q_hL}\bigr)
-\bigl(\cos{\gamma_m^+} - i\sin{\gamma_m^+}\coth{q_eL}\bigr)}
{e^{2i\delta}\bigl( \cos{\gamma_m^-} \!+\! i\sin{\gamma_m^-}\coth{q_hL} \bigr)
\bigl( \cos{\gamma_m^+} \!+ \!i\sin{\gamma_m^+}\coth{q_eL} \bigr)\!-\!1
}\,, \nonumber \\ 
r_{he}^{\downarrow\uparrow}(E)&=&
\frac{-\sin{\gamma_m^+}\sin{\gamma_m^-}e^{i(k_2-k_1)L}e^{-2ik_mL}e^{i\delta}}{
\sinh{q_eL}\sinh{q_hL}\bigl[ 
e^{2i\delta}\bigl( \cos{\gamma_m^-}\!+\! i\sin{\gamma_m^-}\coth{q_hL} \bigr)
\bigl( \cos{\gamma_m^+} \!+\! i\sin{\gamma_m^+}\coth{q_eL} \bigr)\!-\!1 \bigr]
}\,. \label{r-app}
\end{eqnarray}

Similarly, for the case $A_{e}^i=0$ we obtain

\begin{align}
r_{eh}^{\downarrow\uparrow}(E) &= 
\frac{-\sin{\gamma_m^+}\sin{\gamma_m^-}e^{i(k_2-k_1)L}e^{2ik_mL}e^{i\delta}}
{\sinh{q_eL}\sinh{q_hL}\bigl[ 
e^{2i\delta}\bigl( \cos{\gamma_m^-} \!+\! i\sin{\gamma_m^-}\coth{q_hL} \bigr)
\bigl( \cos{\gamma_m^+} \!+\! i\sin{\gamma_m^+}\coth{q_eL} \bigr)\!-\!1\bigr]
}\,, \nonumber \\
r_{hh}^{\downarrow\uparrow}(E) &= e^{-i\phi}e^{2ik_2L}
\frac{
e^{2i\delta}\bigl(\cos{\gamma_m^+} + i\sin{\gamma_m^+}\coth{q_eL}\bigr)
-\bigl(\cos{\gamma_m^-} - i\sin{\gamma_m^-}\coth{q_hL}\bigr)}
{e^{2i\delta}\bigl( \cos{\gamma_m^-} \!+\! i\sin{\gamma_m^-}\coth{q_hL} \bigr)
\bigl( \cos{\gamma_m^+} \!+\! i\sin{\gamma_m^+}\coth{q_eL} \bigr)\!-\!1}\,,
\end{align}
where $r_{eh}^{\downarrow\uparrow}(E) = A_e^o/A_h^i$ and $r_{hh}^{\downarrow\uparrow}(E) = A_h^o/A_h^i$. In these expressions $L$ is the length of the region II.

Finally, after obtaining all the reflection coefficients the 4x4 scattering matrix of the device reads,
\begin{equation}
    S(E) = \left(
    \begin{array}{cccc}
    0 & 0 & 0 & 0 \\
    r_{ee}^{\downarrow\uparrow}(E) & 0 & 0 & r_{eh}^{\downarrow\uparrow}(E) \\
    r_{he}^{\downarrow\uparrow}(E) & 0 & 0 & r_{hh}^{\downarrow\uparrow}(E) \\
    0 & 0 & 0 & 0
    \end{array}
    \right)\,.
\end{equation}

The scattering matrix obeys a particle-hole symmetry
\begin{equation}
\tau_y \sigma_y S(-E)^* \tau_y \sigma_y=S(E),
\end{equation}
which can be written as 
\begin{equation}
r_{ee}^{\downarrow\uparrow}(E)=-[r_{hh}^{\downarrow\uparrow}(-E)]^*, \ r_{eh}^{\downarrow\uparrow}(E)=-[r_{he}^{\downarrow\uparrow}(-E)]^*.
\end{equation}
Moreover, the reflection coefficients satisfy
\begin{equation}
|r_{ee}^{\downarrow\uparrow}(E)|^2+|r_{he}^{\downarrow\uparrow}(E)|^2=1.
\end{equation}

\subsection{Limiting cases of the reflection coefficient}

The expression for the reflection coefficient can be considerably simplified in the parameter regime of interest for our work. We start from the electron-electron reflection coefficient written in a slightly modified way,

\begin{equation}
r_{ee}^{\downarrow\uparrow}(E) = e^{i\phi}e^{-2ik_1L}e^{i\delta}\frac{
e^{i\delta}\bigl(\cos{\gamma_m^-} + i\sin{\gamma_m^-}\coth{q_hL}\bigr)
-e^{-i\delta}\bigl(\cos{\gamma_m^+} - i\sin{\gamma_m^+}\coth{q_eL}\bigr)}
{e^{2i\delta}\bigl( \cos{\gamma_m^-} \!+\! i\sin{\gamma_m^-}\coth{q_hL} \bigr)
\bigl( \cos{\gamma_m^+} \!+ \!i\sin{\gamma_m^+}\coth{q_eL} \bigr)\!-\!1
}\,.
\label{ree_symm}
\end{equation}

Assuming that the energy is much smaller than the gaps induced by the nearby superconductor and ferromagnet, i.e. $E \ll \Delta_0, m_0\sin{\theta}$, the quantities appearing in Eq. (\ref{ree_symm}) are approximated as follows,

\begin{equation}
e^{i\delta} =  \frac{E}{\Delta_0} + i \frac{\sqrt{\Delta_0^2-E^2}}{\Delta_0} 
\approx \frac{E}{\Delta_0} + i\,,
\end{equation}
\begin{equation}
\sin{\gamma_m^{\pm}} =  \frac{\sqrt{m_0^2\sin^2{\theta}-(E\pm eV_g)^2}}{m_0\sin{\theta}} 
\approx  \sqrt{1-\frac{e^2V_g^2}{m_0^2\sin^2\theta}} \mp \frac{eV_g E/m_0\sin{\theta}}{\sqrt{m_0^2\sin^2{\theta}-e^2V_g^2}}\,,
\end{equation}
\noindent and
\begin{equation}
q_{e,h} = \frac{\sqrt{ m_0^2\sin^2\theta - (E\pm eV_g)^2}}{\hbar v_F}
 \approx\frac{1}{\hbar v_F}\left(\sqrt{ m_0^2\sin^2\theta - e^2V_g^2}
\mp \frac{eV_g E}{\sqrt{ {m_0^2\sin^2\theta - e^2V_g^2}}}\right)\,.
\end{equation}

Moreover, in the limit $q_{e,h}L \gg 1$, we can use the asymptotic expressions 
\begin{equation}
\coth{q_{e,h}L} \approx 1 + 2e^{-2q_{e,h}L}\,.
\end{equation}

\begin{figure}
    \centering
    \includegraphics[width=0.8\linewidth]{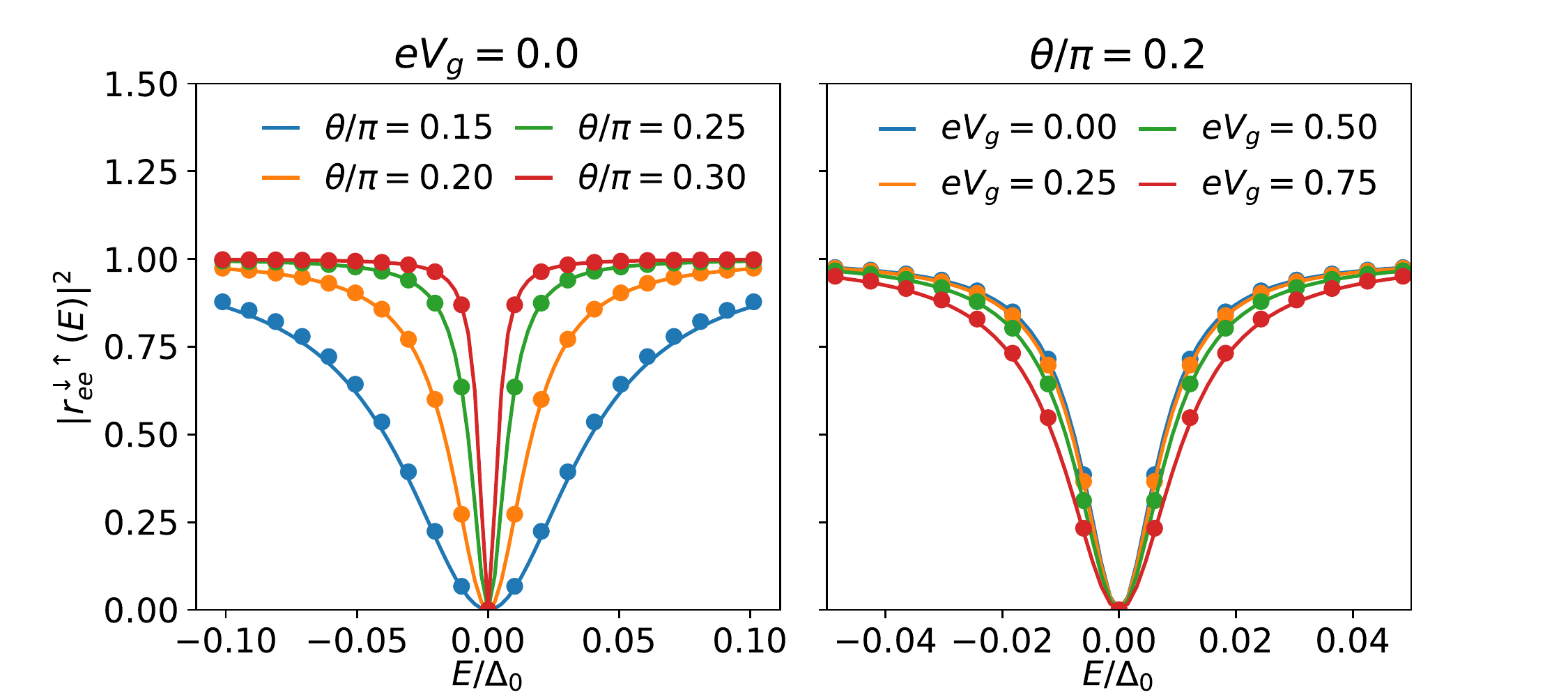}
    \caption{Comparison between the reflection coefficient [Eq.~(\ref{r-app})] (straight lines) and its low energy approximation [Eqs.~(\ref{ree_EandG}) and (\ref{formGamma})] (circles). (a) $|r_{ee}^{\downarrow\uparrow}|^2$ as a function of $E$ for $V_g=0$ and $\theta/\pi=:0.15, 0.20, 0.25, 0.3$. (b) Same for $\theta/\pi=0.2$ and $eV_g/\Delta_0=:0, 0.25, 0.5, 0.75$. The other parameters are  $L=320$ nm, $\Delta_0 = 1$ meV, $m_0/\Delta_0 = 2$, and $v_F = 2.7\times 10^5$ m/s.} \label{ree_compar}
\end{figure}

After straightforward calculations where only the leading order terms in energy are considered in Eq.~(\ref{ree_symm}), we obtain
\begin{equation}
    |r_{ee}^{\downarrow\uparrow}(E)|^2 = \frac{E^2/\Gamma^2}{1 + E^2/\Gamma^2}, \ |r_{he}^{\downarrow\uparrow}(E)|^2=\frac{1}{1 + E^2/\Gamma^2},\,
    \label{ree_EandG}
\end{equation}
where 
\begin{equation}
    \Gamma = 2\Delta_0\left(\frac{\xi_F(0,\theta)}{\xi_F(V_g,\theta)}\right)^2
    \frac{\xi_S}{\xi_F(V_g,\theta)+\xi_S}
    e^{-2L/\xi_{F}(V_g,\theta)}\, 
\label{formGamma}
\end{equation}
is the Majorana linewidth (for which $|r_{ee}^{\downarrow\uparrow}(E=\Gamma)|^2=|r_{he}^{\downarrow\uparrow}(E)|^2=1/2$) expressed in terms of the coherence lengths
$\xi_F(V_g,\theta)=\hbar v_F/{\sqrt{m_0^2\sin^2\theta-(eV_g)^2}}$ and $\xi_S=\hbar v_F/\Delta_0$ pertaining to the ferromagnet and superconductor, respectively. Figure~\ref{ree_compar} shows that the comparison of the approximate expressions for the scattering coefficients [Eqs.~(\ref{ree_EandG}) and (\ref{formGamma})] are in excellent agreement with the full expression
[Eq.~(\ref{r-app})].

\section{Asymptotic expressions for the dimensionless charge $\mathcal{Q}$}

As discussed in the main text for circular precession of the magnetization around the $z$-axis the pumped charge, spin and heat during one cycle are given by
\begin{equation}
Q_e = e\mathcal{Q}, \quad S_z = -\frac{\hbar}{2}\mathcal{Q}, \quad Q_E = \frac{\hbar\omega}{2}\mathcal{Q}\,,
\label{universal-app}
\end{equation}
with 
\begin{equation}
\mathcal{Q} =-\frac{1}{2\pi}\! \int\!dE \frac{\partial f(E-eV_d)}{\partial E}
\int_0^{2\pi} \!d\phi\,\bigl |r_{ee}^{\downarrow\uparrow}(E,\phi)\bigr|^2=- \int\!dE \frac{\partial f(E-eV_d)}{\partial E}
\bigl |r_{ee}^{\downarrow\uparrow}(E)\bigr|^2\,,
\label{C2}
\end{equation}
where we have used that $|r_{ee}^{\downarrow\uparrow}(E)\bigr|$ does not depend on $\phi$.

At $T=0$, we obtain 
\begin{equation}
\mathcal{Q} =  \bigl|r_{ee}^{\downarrow\uparrow}(eV_d)\bigr|^2\,,
\end{equation}
so that we straightforwardly  obtain
\begin{equation}
    \mathcal{Q} \approx \left\{
    \begin{array}{lc}
     \displaystyle{ \frac{e^2V_d^2}{\Gamma^2}}\,,     & \quad eV_d\ll \Gamma     \\\\
     \displaystyle{1-\frac{\Gamma^2}{e^2V_d^2}}\,,   & \quad eV_d\gg \Gamma     
    \end{array}
    \right.\,.
\end{equation}

On the other hand, at $eV_d=0$ we obtain
\begin{equation}
\mathcal{Q} = \int\!dE \frac{\beta e^{\beta E}}{(e^{\beta E}+1)^2}
\bigl |r_{ee}^{\downarrow\uparrow}(E)\bigr|^2=\int\!dE \frac{\beta e^{\beta E}}{(e^{\beta E}+1)^2} \frac{E^2/\Gamma^2}{1+E^2/\Gamma^2}\,,
\end{equation}
where $\beta=1/k_B T$. At the low and high temperature limits we obtain
\begin{equation}
    \mathcal{Q} \approx \left\{
    \begin{array}{lc}
     \displaystyle{ 
     \!\left(\frac{k_BT}{\Gamma}\right)^2\int_{-\infty}^\infty\! dx \frac{x^2 e^x}{(e^x +1)^2}}=\frac{\pi^2}{3}\left(\frac{k_BT}{\Gamma}\right)^2\,,     & \quad k_BT \ll \Gamma     \\\\
     \displaystyle{
     1
     -\frac{\Gamma}{4k_BT}\!\int_{-\infty}^{\infty}\! dx \frac{1}{1+x^2}=1-\frac{\pi}{4}\left(\frac{\Gamma}{k_BT}\right)\,,}   & \quad k_B T \gg \Gamma     
    \end{array}
    \right.\,.
\end{equation}

\section{Comparison with numerical simulations}

In this section numerical simulations of the device introduced in the main text are presented using the tight-binding method. The time-dependent Bogoliubov-de Gennes (BdG) Hamiltonian describing the QSHI in proximity contact with superconductivity and ferromagnetism reads
\begin{equation}
    \mathcal{H}_{BdG}(t) = [\mathcal{H}({\bs k}) - \mu({\bs x})]\tau_z + {\bs m}({\bs x},t)\cdot{\bs \sigma} + \Delta({\bs x})\tau_x\,, 
\label{2DtBdG}
\end{equation}

\noindent where ${\bs \sigma}=(\sigma_x,\sigma_y,\sigma_z)$ and ${\bs \tau}=(\tau_x,\tau_y,\tau_z)$ are Pauli matrices acting in the spin and Nambu space respectively, ${\bs m}({\bs x},t)=m_0({\bs x})\left(\sin{\theta\cos\omega t},\sin{\theta}\sin{\omega t},\cos{\theta}\right)$ is the circularly precessing magnetization in the FI, $\Delta({\bs x})$ is the superconducting order parameter, and $\mu({\bs x})$ is the chemical potential. To describe the QSHI we consider Bernevig-Hughes-Zhang (BHZ) Hamiltonian 
\begin{equation}
\mathcal{H}({\bs k}) = 
Aa^{-1}\sin{(k_x a)}s_x\sigma_z -Aa^{-1}\sin{(k_y a)}s_y \sigma_0 + 
\bigl(M-Ba^{-2}[4 - 2\cos{k_x a} - 2\cos{k_y a}]\bigr)s_z \sigma_0 \,,
\end{equation}

\noindent where ${\bs s}=(s_x,s_y,s_z)$ are Pauli matrices acting on the orbital states and $a$ is the lattice constant. The  Hamiltonian of Eq.~(\ref{2DtBdG}) is simulated for a slab of width $W=$ 220 nm where the magnetization is introduced over a region of width $L_y$ on one of the edges. More explicitly, we use 
\begin{eqnarray}
    m_0({\bs x})
    &=& 
     \left\{
    \begin{array}{cc}
        \displaystyle{m_0},  &  \quad 0 \leq x \leq L_x \quad \mathrm{and}\quad 0 \leq y \leq L_y, \\
        \displaystyle{0},  & \quad \displaystyle{\mathrm{elsewhere}},
    \end{array}
    \right. \\
    \Delta({\bs x})
    &=& 
     \left\{
    \begin{array}{cc}
        \displaystyle{\Delta_0},  &  \quad x > L_x, \\
        \displaystyle{0},  & \quad \displaystyle{\mathrm{elsewhere}},
    \end{array}
    \right.\\
    \mu({\bs x})
    &=& 
     \left\{
    \begin{array}{cc}
        \displaystyle{eV_g},  &  \quad 0 \leq x \leq L_x \quad \mathrm{and}\quad 0 \leq y \leq L_y,\\
        \displaystyle{0},  & \displaystyle{\quad x < 0 },\\
        \displaystyle{0},  & \quad \displaystyle{\mathrm{elsewhere}}.
    \end{array}
    \right.
\end{eqnarray}
We assume that the size of the FI region is $L_x =$ 340 nm and $L_y=$ 66 nm, and the rest of the parameters defining the Hamiltonian are listed on Table \ref{paramtable}. 

\begin{table}[h]
\begin{tabularx}{0.75\textwidth}{|>{\centering\arraybackslash}X |>{\centering\arraybackslash}X |>{\centering\arraybackslash}X |>{\centering\arraybackslash}X |>{\centering\arraybackslash}X 
|>{\centering\arraybackslash}X|}
\hline
$A$ [meV \!$\cdot$\! nm] & $B$ [meV \!$\cdot$\! nm$^2$] & $M$ [meV] & $m_0$ [meV] & $\Delta_0$ [meV] & $a$ [nm]\\ 
\hline
180 & 340 & 10 & 2 & 1 & 4 \\
\hline
\end{tabularx}
\caption{Parameters used in the numerical simulations of the time-dependent Bogoliubov-de Gennes Hamiltonian of Eq.~(\ref{2DtBdG}).}
\label{paramtable}
\end{table}

\begin{figure}[htb]
    \centering
    \includegraphics[width=0.95\linewidth]{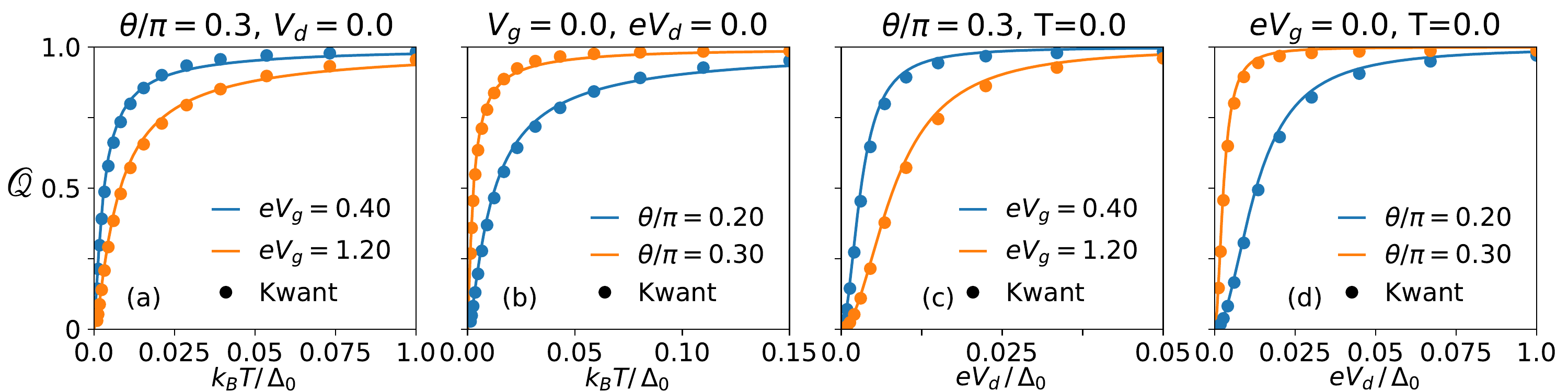}
    \caption{The dimensionless charge $\mathcal{Q}$ as a function of temperature (a-b) and draining voltage (c-d) from both full lattice diagonalization (dots) and from the effective one-dimensional model [Eqs.~(\ref{C2})] (lines). Here, $L_=340$ nm, $\Delta_0 = 1$ meV, $m_0/\Delta_0 = 2$, and $v_F = 2.7\times 10^5$ m/s. The parameters of the lattice model are described in Table \ref{paramtable} and the width of the FI region is $L_y=$ 66 nm.} 
    \label{QvsT_1d&lat}
\end{figure}

The scattering matrix of our device is calculated using the Kwant package. The electric charge $Q_e$, spin $S_z$ and heat $Q_E$ pumped over a cycle are calculated from the scattering matrix as discussed in the main text. In the lattice case we also obtain the same  dependence of $Q_e$, $S_z$ and $Q_E$ on the dimensionless charge $\mathcal{Q}$ as it was demonstrated in the main text within the effective one-dimensional model. In Fig.~\ref{QvsT_1d&lat} a comparison between the one-dimensional (lines) and lattice (dots) cases reveals excellent agreement between the two methods. The dimensionless charge is here plotted as a function of temperature (a-b) and draining voltage (c-d) for various values of the polar angle $\theta$ and the gate voltage $V_g$. 

\section{Robustness against disorder}

\begin{figure}[htb]
    \centering
    \includegraphics[width=0.95\linewidth]{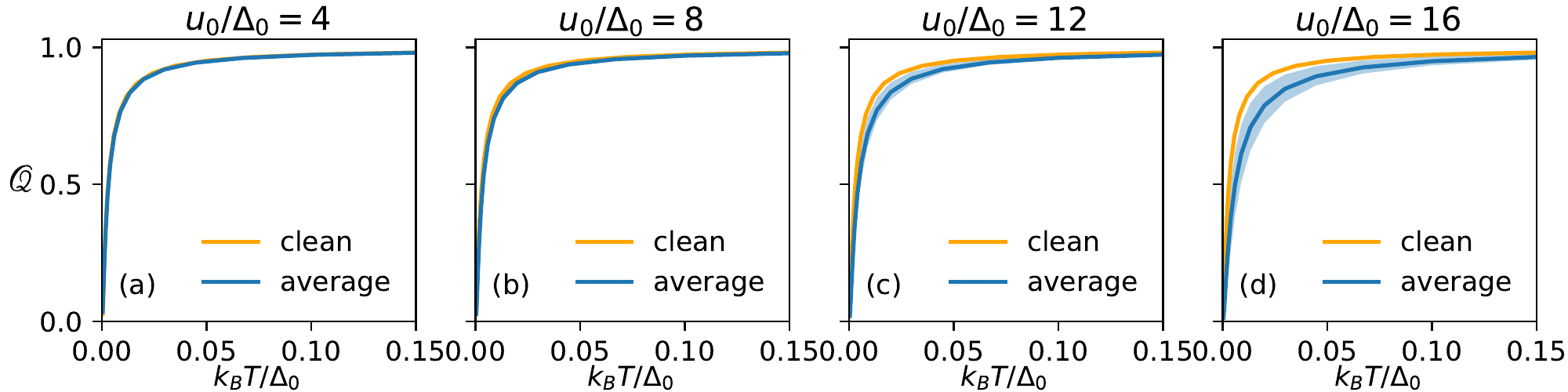}
    \caption{Average over several disorder configurations of the dimensionless charge $\mathcal{Q}$, plotted as a function of temperature for four different disorder strengths, $u_0/\Delta_0=4,\,8,\,12,$ and $16$. The shade region around the average charge shows the 80\% percentile error. For comparison the disorder-free (clean) dimensionless charge is also shown in the figures. Here, $\theta/\pi=0.3$, $V_g=V_d=0$, and the parameters of the lattice model are described in Table \ref{paramtable}.  } 
    \label{QvsT_disor}
\end{figure}

In this section we calculate the robustness of the pumped quantities in the two operation regimes, i.e. the regime where the pumped quantities are quantized and the regime where they are suppressed. Disorder is included in hamiltonian~\ref{2DtBdG} by introducing a random potential in the normal hamiltonian $\mathcal{H}(\mathbf{k})$. The disorder potential reads

\begin{equation}
    u({\bs x}) = u_0 \mathrm{O}({\bs x}),
\end{equation}

\noindent where $\mathrm{O}({\bs x})$ is a random spatial distribution of numbers between 1 and -1 over the scattering region $0 \leq x \leq L_x$ and $0 \leq y \leq L_y$. $u_0$ is the strength of the disorder and we consider four different values,
$u_0/\Delta_0=4,\,8,\,12,$ and $16$. The effect of disorder on the pumped quantities is analyzed statistically by considering one hundred different random spatial distributions. Fig.~\ref{QvsT_disor} summarizes the results of the statistical analysis.

The average of the dimensionless charge $\mathcal{Q}$ over all the random distributions are plotted in Fig.~\ref{QvsT_disor} along with the disorder-free (clean) $\mathcal{Q}$. The error over the disorder average is also shown in the figure. The error was obtained from the bonds of the 80\% percentile deviations of the average value. The error that becomes small at the operation regimes  $T\approx 0$ and $T\gg \Gamma$, where $\Gamma$ is the Majorana linewidth, demonstrates topological protection in these two regimes.    

\section{Rotating wave description of the scattering}

Starting from the time-dependent Hamiltonian
\begin{equation}
    \mathcal{H}(t)=[v_Fp\,\sigma_z-\mu(x)]\tau_z+{\bs m}(x,t)\cdot{\bs \sigma}+\Delta(x)\tau_x\,, 
\end{equation}
where $m_z(x)=m_0(x) \cos [\theta(x)]$, $m_x(x, t)=m_0(x) \sin [\theta(x)] \cos \omega t$ and $m_y(x, t)=m_0(x) \sin [\theta(x)] \sin \omega t$, 
we can perform a unitary transformation $U=e^{-i\omega t \sigma_z/2}$, which brings the Hamiltonian in the time-independent form
\begin{equation}
H_{\rm eff}=U^\dag H(t) U-i\hbar U^\dag \frac{\partial U}{\partial t}=[v_Fp\,\sigma_z-\mu(x)]\tau_z+m_0(x) \cos [\theta(x)] \sigma_z+m_0(x) \sin [\theta(x)] \sigma_x +\Delta(x)\tau_x -\frac{\hbar \omega}{2} \sigma_z\,, \label{effective-Hamiltonian}
\end{equation}
but there now exists a fictitious spin bias term $-\frac{\hbar \omega}{2} \sigma_z$.  Due to the non-equilibrium nature of the problem this term acts not only on the spectrum but it also shifts the distribution functions of the spin up and spin down electrons in leads. Otherwise, the transport in this system system can be calculated in similar way as in the case of static Hamiltonians.

In particular, the reflection coefficients should be calculated from Eq.~(\ref{effective-Hamiltonian}). It is straightforward to see that the Majorana zero mode is pinned to zero energy also in the presence of the term $-\frac{\hbar \omega}{2} \sigma_z$. Thus, this term can only make the Majorana linewidth $\Gamma(\omega)$ frequency dependent, and we obtain 
\begin{equation}
    |r_{ee}^{\downarrow\uparrow}(E)|^2 = \frac{E^2/\Gamma^2(\omega)}{1 + E^2/\Gamma^2(\omega)}\,,\, \ |r_{he}^{\downarrow\uparrow}(E)|^2 = \frac{1}{1 + E^2/\Gamma^2(\omega)}\,,
    \label{ree_EandGomega}
\end{equation}
where $\Gamma(\omega)$ is the same as in Eq. (\ref{formGamma}) but with $\xi_S$ now being
\begin{equation}
\xi_S(\omega)=\frac{\hbar v_F}{\sqrt{\Delta_0^2-(\hbar\omega/2)^2}}\,.   
\end{equation}
\noindent From this expression, we find that the frequency dependence of $\Gamma(\omega)$ arises only from an effective renormalization of the superconducting gap, and as long as $\hbar \omega \ll \Delta_0$ the linewidth $\Gamma(\omega)$ depends only weakly on the frequency. Due to this reason, in the rest of the section we neglect the frequency-dependence of $\Gamma(\omega)$. Throughout this section we assume that  $\hbar \omega \ll \Delta_0, m_0 \sin \theta$ are satisfied.

\subsection{Charge current}

By including all the scattering paths in the presence of true and fictitious biases the charge current flowing into the normal lead can be written as 
\begin{align}
\langle I_e \rangle &= \frac{e}{h} \int dE \bigg[f\left(E-\frac{\hbar \omega}{2} - eV_d\right)-f\left(E+\frac{\hbar \omega}{2}\right)\bigg]
|r_{ee}^{\downarrow\uparrow}(E)|^2\nonumber\\
&-\frac{e}{h} \int dE \bigg[f\left(E-\frac{\hbar \omega}{2} - eV_d\right)-f\left(E-\frac{\hbar \omega}{2}\right)\bigg]
\bigg(|r_{he}^{\downarrow\uparrow}(E)|^2+1\bigg)\,.
\end{align}

\subsubsection{Low-frequency limit}
Assuming that  $\hbar \omega \ll k_B T$ or $\hbar\omega \ll \Gamma$ is satisfied, we obtain after a straightforward calculation that the charge current can be expressed as
\begin{equation}
\langle I_e \rangle =I_e^{\rm dc}(eV_d)+I_e^{\rm pump}\,.
\end{equation}
Here $I_e^{\rm dc}(eV_{d})$ is the standard dc current-voltage characteristics originating from the Andreev reflection caused by the Majorana zero mode. It has negative sign because the current is flowing away from the lead. Moreover, it is independent on the magnetization dynamics and it is given by expression
\begin{equation}
I_e^{\rm dc}(eV_d)=- \frac{2e}{h} \int dE \bigg[f\big(E - eV_d\big)-f\big(E\big)\bigg] |r_{he}^{\downarrow\uparrow}(E)|^2\,.
\end{equation}
On the other hand, the charge current caused by the rotating magnetization is given by 
\begin{equation}
I_e^{\rm pump}=-\frac{e \omega}{2 \pi} \int dE \frac{\partial f(E-eV_d)}{\partial E} |r_{ee}^{\downarrow\uparrow}(E)|^2\,.
\end{equation}
This agrees with the expression given in the main text for the charge current.

\subsubsection{Frequency dependence}

If $\hbar \omega \gtrsim k_B T$ and $\hbar\omega \gtrsim \Gamma$ there can be strong frequency-dependent corrections to these results. To illustrate this we compute the charge current for $V_d=0$ and $T=0$ 
\begin{equation}
\langle I_e \rangle = \frac{e}{h} \int^{\hbar \omega/2}_{-\hbar \omega/2} dE 
\frac{E^2/\Gamma^2}{1 + E^2/\Gamma^2}=\frac{2e \Gamma}{h} \bigg[\frac{\hbar \omega}{2\Gamma}-\arctan\bigg(\frac{\hbar \omega}{2 \Gamma}\bigg) \bigg]\,.
\end{equation}
In the different limits we obtain 
\begin{equation}
\langle I_e \rangle = \begin{cases} \displaystyle{\frac{e \omega}{2\pi} \frac{\hbar^2 \omega^2}{12 \Gamma^2}}\,, & |\hbar \omega| \ll \Gamma \\\\
\displaystyle{\frac{e \omega }{2 \pi} \left(1- \frac{\pi \Gamma}{ \hbar |\omega|} \right)}\,, & |\hbar \omega| \gg \Gamma\,
\end{cases}
\,.
\end{equation}
In the low-frequency limit $\hbar \omega \ll \Gamma$ the current is suppressed, whereas in the limit $\hbar \omega \gg \Gamma$ it approaches the quantized value $\langle I_e \rangle = e \omega/2\pi$.

\subsubsection{Comparison of the dc and pumped current}

In the regime of interest $eV_d, \hbar\omega, eV_g, k_BT \ll \Delta_0, m_0 \sin \theta$, the dc charge current is always bounded by the expression 
\begin{equation}
|I_e^{\rm dc}|<\frac{2e}{h} \int_{-\infty}^\infty dE \  |r^{eh}_{\uparrow\downarrow}(E)|^2= e\frac{\Gamma}{\hbar}\,.
\end{equation}

By expressing the pumped current as $I_e^{\rm pump} =eQ\omega/2 \pi$, we see that the pumped charge current is  larger than the dc charge current if 
\begin{equation}
Q_e > \frac{2\pi \Gamma}{\hbar \omega}\,.
\end{equation}
Thus the dc current can be neglected if $\hbar \omega \gg \Gamma$. Moreover, the dc current can also be neglected if $eV_d \ll \hbar \omega$. Alternatively the pumped charge current can be measured by taking the difference of the currents measured in the presence and absence of the magnetization precession.

\subsubsection{Differential conductance resonance at half-integer frequency}

At $T=0$ the differential conductance obeys
\begin{equation}
\frac{d \langle I_e \rangle}{d V_d}=-\frac{2e^2}{h} \frac{1}{\displaystyle{1+\left(\frac{\hbar \omega}{2 \Gamma}+\frac{eV_d}{\Gamma}\right)^2}}\,. 
\end{equation}
Thus, there is a peak in the differential conductance at $eV_d=-\hbar\omega/2$. 
This peak arises at fractional frequency because Majorana zero mode is a spinless excitation appearing at energy $E=0$. Due to this reason, the fictitious spin bias in Eq.~(\ref{effective-Hamiltonian}) does not influence the energy of the Majorana zero mode, so that $r_{he}^{\downarrow\uparrow}(E)$ is frequency independent (apart from the weak frequency dependence of the linewidth). This can be contrasted to the case where the excitation would have a spin so that $r_{he}^{\downarrow\uparrow}(E)= r_{0,he}^{\downarrow\uparrow}(E \pm \hbar \omega/2)$, where the Andreev reflection coefficient in the absence of the spin bias $r_{0,he}^{\downarrow\uparrow}(E)$ has a resonance at energy of the excitation $E=E_0$. In the latter case, there is a peak in the differential conductance at $eV_d=E_0+n \hbar \omega$ ($n=0,-1$). Indeed, robust resonances in the presence of time-dependent driving are generically expected to appear at voltages $eV_d$ which are integer multiples of $\hbar \omega$, and therefore the robust peak appearing at half-integer frequency is a peculiar consequence of the properties of the Majorana zero mode.

\subsection{Spin current}

Similarly, the spin current can be expressed as 
\begin{align}
\langle I_s \rangle &= -\frac{\hbar}{2} \frac{1}{h} \int dE \bigg[f\left(E-\frac{\hbar \omega}{2} - eV_d\right)-f\left(E+\frac{\hbar \omega}{2}\right)\bigg]
|r_{ee}^{\downarrow\uparrow}(E)|^2\nonumber\\
&+\frac{\hbar}{2}\frac{1}{h} \int dE \bigg[f\left(E-\frac{\hbar \omega}{2} - eV_d\right)-f\left(E-\frac{\hbar \omega}{2}\right)\bigg]
\bigg(|r_{he}^{\downarrow\uparrow}(E)|^2-1\bigg)\,. \label{spin-cur-rot-frame}
\end{align}

\subsubsection{Low-frequency limit}

Assuming that  $\hbar \omega \ll k_B T$ or $\hbar\omega \ll \Gamma$ is satisfied, we obtain after a straightforward calculation that the spin current can be expressed as
\begin{equation}
\langle I_s \rangle =I_s^{\rm dc}(eV_d)+I_s^{\rm pump}\,.
\end{equation}
Here $I_s^{\rm dc}(eV_{d})$ is the standard spin current-voltage characteristics originating from the spin reflection caused by the ferromagnet (each spin reflection results in an injection of a spin $\pm 2 \times \hbar/2$ into the ferromagnet). It is independent on the magnetization dynamics and it is given by expression
\begin{equation}
I_s^{\rm dc}(eV_d)=-2 \frac{\hbar}{2} \frac{1}{h} \int dE \bigg[f\big(E - eV_d\big)-f\big(E\big)\bigg] |r_{ee}^{\downarrow\uparrow}(E)|^2\,.
\end{equation}
On the other hand, the spin current caused by the rotating magnetization is given by 
\begin{equation}
I_s^{\rm pump}=\frac{\hbar}{2}\frac{\omega}{2 \pi} \int dE \frac{\partial f(E-eV_d)}{\partial E} |r_{ee}^{\downarrow\uparrow}(E)|^2\,.
\end{equation}
This agrees with the expression given in the main text for the spin current.

\subsubsection{Frequency dependence}

We can also calculate the frequency dependent corrections in the case $\hbar \omega \gtrsim k_B T$ and $\hbar\omega \gtrsim \Gamma$. If $V_d=0$ and $T=0$, we obtain from Eq.~(\ref{spin-cur-rot-frame})
\begin{equation}
\langle I_s \rangle = - \frac{\hbar}{2}\frac{2\Gamma}{h} \bigg[\frac{\hbar \omega}{2\Gamma}-\arctan\bigg(\frac{\hbar \omega}{2 \Gamma}\bigg) \bigg]\,.
\end{equation}
In the different limits, we obtain 
\begin{equation}
\langle I_s \rangle = \begin{cases} \displaystyle{-\frac{\hbar}{2}\frac{\omega}{2\pi} \frac{\hbar^2 \omega^2}{12 \Gamma^2}}\,, & |\hbar \omega| \ll \Gamma \\\\
\displaystyle{-\frac{\hbar}{2}\frac{\omega }{2 \pi} \left(1- \frac{\pi \Gamma}{ \hbar |\omega|} \right)}\,, & |\hbar \omega| \gg \Gamma
\end{cases}
\,.
\end{equation}
The expressions are similar as for the charge current except that $e$ is replaced by $-\hbar/2$.

\subsubsection{Comparison of the dc and pumped spin currents}

There is an important difference between the charge and spin currents. Namely, the dc spin current is not bounded similarly as the dc charge current. Instead, for the dc spin current we obtain for $eV_d \gg \Gamma$ or $k_BT \gg \Gamma$ a bound
\begin{equation}
|I_s^{\rm dc}(eV_d)|<  \frac{\hbar}{2} \frac{e V_d}{\pi \hbar}\,.
\end{equation}
On the other hand, if $k_B T=0$ we obtain
\begin{equation}
|I_s^{\rm dc}(eV_d)|=2 \frac{\hbar}{2} \frac{\Gamma}{h} \bigg[\frac{eV_d}{\Gamma}-\arctan\left(\frac{eV_d}{\Gamma}\right) \bigg]\,.
\end{equation}

Therefore the pumped current dominates the dc current if $\hbar \omega \gg eV_d$. Alternatively the pumped spin current can be measured by taking the difference of the currents measured in the presence and absence of the magnetization precession. 

\subsection{Heat current}

Similarly, the heat current can be expressed as
\begin{eqnarray}
\langle I_E \rangle &=& \frac{1}{h} \int dE\left(E+\frac{\hbar \omega}{2}\right) \bigg[f\left(E-\frac{\hbar \omega}{2} - eV_d\right)-f\left(E+\frac{\hbar \omega}{2}\right)\bigg]
|r_{ee}^{\downarrow\uparrow}(E)|^2 \nonumber\\
&&+\frac{1}{h} \int dE \left(E-\frac{\hbar \omega}{2}\right) \bigg[f\left(E-\frac{\hbar \omega}{2} - eV_d\right)-f\left(E-\frac{\hbar \omega}{2}\right)\bigg]
\bigg(|r_{he}^{\downarrow\uparrow}(E)|^2-1\bigg)\,. \label{heat-cur-rot-frame}
\end{eqnarray}
Here we have taken into account that the spin-$\downarrow$ electrons  can dispose as heat the energy $E-\mu_\downarrow$, where $\mu_\downarrow=-\hbar\omega/2$ is the effective chemical potential which takes into account the fictitious spin bias. Similarly, the spin down holes can dispose an energy $E+\mu_\downarrow$ as heat in the lead.

\subsubsection{Low-frequency limit}
Assuming that  $\hbar \omega \ll k_B T$ or $\hbar\omega \ll \Gamma$ is satisfied, we obtain after a straightforward calculation that the heat current can be expressed as
\begin{equation}
\langle I_E \rangle =I_E^{\rm dc}(eV_d)+I_E^{\rm pump}.
\end{equation}
Here, the dc contribution to the heat current arises via the voltage-driven spin flips that inject or absorb magnons from the ferromagnetic insulator
\begin{equation}
I_E^{\rm dc}(eV_d) = \hbar \omega \frac{I_s^{\rm dc}(eV_d)}{\hbar}=\hbar \omega \frac{1}{h} \int dE  \bigg[f\big(E - eV_d\big)-f\big(E\big)\bigg]
|r_{ee}^{\downarrow\uparrow}(E)|^2.
\end{equation}
and the heat current caused by the precessing magnetization is given by 
\begin{equation}
I_E^{\rm pump} = - \frac{\hbar \omega}{2} \frac{\omega}{2\pi} \int dE  \frac{\partial f(E-eV_d)}{\partial E} 
|r_{ee}^{\downarrow\uparrow}(E)|^2. 
\end{equation}
The latter formula agrees with the expression given in the main text for the heat current.

\subsubsection{Frequency dependence}

We can also calculate the frequency dependent corrections in the case $\hbar \omega \gtrsim k_B T$ and $\hbar\omega \gtrsim \Gamma$. If $V_d=0$ and $T=0$, we obtain from Eq.~(\ref{heat-cur-rot-frame})
\begin{eqnarray}
\langle I_E \rangle &=& \frac{\hbar \omega}{2} \frac{1}{h} \int_{-\hbar \omega/2}^{\hbar \omega/2} dE  
|r_{ee}^{\downarrow\uparrow}(E)|^2= \frac{\hbar\omega}{2}\frac{2\Gamma}{h} \bigg[\frac{\hbar \omega}{2\Gamma}-\arctan\bigg(\frac{\hbar \omega}{2 \Gamma}\bigg) \bigg]\,.
\end{eqnarray}
In the different limits we obtain 
\begin{equation}
\langle I_E \rangle = \begin{cases} \displaystyle{\frac{\hbar \omega}{2}\frac{\omega}{2\pi} \frac{\hbar^2 \omega^2}{12 \Gamma^2}}\,, & |\hbar \omega| \ll \Gamma \\\\
\displaystyle{\frac{\hbar \omega}{2}\frac{\omega }{2 \pi} \left(1- \frac{\pi \Gamma}{ \hbar |\omega|} \right)}\,, & |\hbar \omega| \gg \Gamma
\end{cases}
\,.
\end{equation}
The expressions are similar as for the spin current except that in the prefactor $-\hbar/2$ is replaced by $\hbar \omega/2$.

\subsubsection{Comparison of the dc and pumped heat currents}

The dc and pumped heat currents are related to spin currents. Therefore, based on the results derived in the previous section, we can conclude that 
the pumped heat current dominates the dc heat current if $\hbar \omega \gg eV_d$.

\section{Thouless-like pumping in a hybrid superconductor-ferromagnet structure}

For completeness, we consider also pumped charge current in the Hamiltonian formalism in the presence of superconductivity.
The system is translationally invariant in the $x$-direction so that momentum $k$ along the edge is a good quantum number but in the $y$-direction we include the superconductivity and rotating magnetization only on one of the edges over a region $L_y$. (To keep the formalism simple we open an energy gap also at the other edge by covering it with a FI that has a static magnetization. Alternatively, the gap could be opened by covering it with superconductivity.) 
The BdG Hamiltonian for the system is
\begin{equation}
    \mathcal{H}_{BdG}(k, t) = [\mathcal{H}({\bs k}) - \mu]\tau_z + {\bs m}(y,t)\cdot{\bs \sigma} + \Delta(y)\tau_x\,, 
\label{2DtBdGNHam}
\end{equation}
where ${\bs m}(y,t)=m_0(y)[\sin{\theta(y)}\cos\phi(y, t),\sin{\theta(y)} \sin{\phi(y, t)},\cos{\theta(y)}]$.
More explicitly, we use 
\begin{equation}
    m_0(y)=
     \begin{cases}
        m_0,  &  0 \leq y \leq L_y, \\
        m_0,  &  W-L_y \leq y \leq W, \\
        0,  & \mathrm{elsewhere},
    \end{cases} \label{m0ylattice}
\end{equation}
\begin{equation}
\theta(y)=
   \begin{cases}
        \theta,  &  0 \leq y \leq L_y, \\
        \theta_2,  &  W-L_y \leq y \leq W, \\
        0,  & \mathrm{elsewhere},
    \end{cases} \label{thetaylattice}
\end{equation}
\begin{equation}
\phi(y)=
   \begin{cases}
        \phi(t),  &  0 \leq y \leq L_y, \\
        \phi_2,  &  W-L_y \leq y \leq W, \\
        0,  & \mathrm{elsewhere},
    \end{cases} \label{phiylattice}
\end{equation}
\begin{equation}
\Delta(y)=
    \begin{cases}
        \Delta_0,  &  0 \leq y \leq L_y, \\
        0,  & \mathrm{elsewhere}.
    \end{cases} \label{Deltaylattice}
\end{equation}    
We assume that $L_y=$ 66 nm, $m_0=2$ meV,  $\phi_2=0$, $\theta_2=0.4 \pi$, $\mu=0$ and the rest of the parameters defining the Hamiltonian are listed on Table \ref{paramtable}. Parameters $\theta$ and $\Delta_0$ are varied to study the effect of superconductivity on the charge pumping. 
\begin{figure}
    \centering
    \includegraphics[width=0.5\linewidth]{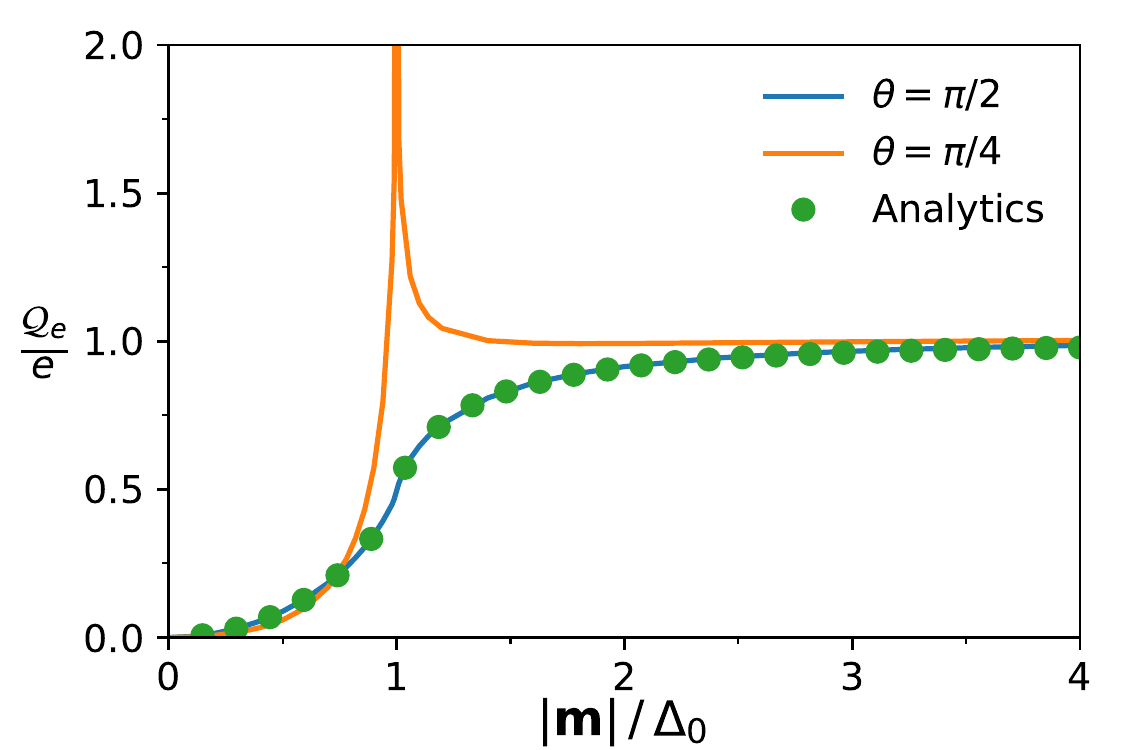}
    \caption{The dimensionless charge $Q_e/e$ as a function of $|\bs m|/\Delta_0$ for various precession angles  both for the full lattice diagonalization (lines) and from the effective one-dimensional model [Eq.~(\ref{analytical_charge})] (circles). The parameters of the lattice model (\ref{2DtBdGNHam}) are described in Eqs.~(\ref{m0ylattice}-\ref{Deltaylattice}) and in Table \ref{paramtable}.} 
    \label{QBerry}
\end{figure}

We can now repeat the calculation of the charge current in the presence of superconductivity. The main difference to the earlier case is that we now the time-dependent current expectation value is 
\begin{equation}
j_c(t)=\frac{1}{2\pi} \int_0^{2\pi} dk~\langle j_c(k,t) \rangle_{s},
\end{equation}
where the current density operator is defined as 
\begin{equation}
j_c(k,t)=\frac{e}{2 \hbar} \partial_{k}\mathcal{H}_{BdG}(k, t) \tau_z
\end{equation}
and the average being taken over the time-dependent state of the system. Therefore, the current expectation value in leading order in velocities is 
\begin{align}
    j_c(t)&=\frac{e}{2h}\sum_{n\in occ} \int_0^{2\pi} dk~\widetilde{\langle n k,t|}\partial_{k}\mathcal{H}_{BdG}(k, t) \tau_z \widetilde{|n k,t\rangle}\nonumber\\
    &+ \frac{e}{2\pi} {\rm Im}\sum_{n\in occ, n'\in empty} \int_0^{2\pi} dk \displaystyle{\frac{\dot{\bs m}(t)\cdot\widetilde{\langle n k,t|}\partial_k \mathcal{H}_{BdG}(k, t) \tau_z \widetilde{|n' k,t\rangle}\widetilde{\langle n' k,t|}\partial_{\bs m}\mathcal{H}_{BdG}(k, t)\widetilde{|n k,t\rangle}}{[\epsilon_n(k, t)-\epsilon_{n'}(k, t)]^2}},
\end{align}
where $\widetilde{|n k, t\rangle}$ are the instantaneous eigenstates of Hamiltonian (\ref{2DtBdGNHam}) in the original (lab) frame. The only time-dependent parameter is $\phi(t)$, and we assume that it goes from $0$ to $2\pi$ during one cycle. The first term describes instantaneous current and it is independent of the time-dependence of $\phi(t)$. (See a more detailed discussion of the first term in Sec.~\ref{insta_sect}) Thus, the second term describes the pumped charge caused by the dynamics of the magnetization
\begin{equation}
    Q_e=e \frac{1}{2\pi} \sum_{n\in occ, n'\in empty} \int_0^{2\pi}d\phi \int_0^{2\pi} dk \ {\rm Im} \bigg\{ \frac{ \widetilde{\langle n k,\theta, \phi|}\partial_k\mathcal{H}_{BdG}(k, \phi) \tau_z\widetilde{|n' k, \theta, \phi\rangle}\widetilde{\langle n' k,\theta, \phi|}\partial_\phi\mathcal{H}_{BdG}(k, \phi)\widetilde{|n k, \theta, \phi\rangle}}{[\epsilon_n(k, \theta, \phi)-\epsilon_{n'}(k, \theta,  \phi)]^2} \bigg\}.
\end{equation}
While naively this expression resembles the Chern number associated with Thouless pumping, which results in the quantization of $Q_e$ in the absence of superconductivity, the presence of the operator $\tau_z$ in  $j_c(t)$ implies that $Q_e$ is not necessarily quantized.  Using Kwant, we have evaluated $Q_e$ numerically as a function of $m_0/\Delta_0$, and we find that 
\begin{equation}
Q_e=\begin{cases}
0 & |\mathbf{m}|/\Delta_0 \to 0 \\
1 & |\mathbf{m}|/\Delta_0 \gg 1.
\end{cases}
\end{equation}
The full dependence of $Q_e$ on $m_0/\Delta_0$ is shown in Fig.~\ref{QBerry}. We have also been able to calculate analytically  $Q_e$  from the effective low-energy theory 
\begin{align}
    H_{edge}(p, t)&=v_Fp\,\sigma_z\tau_z+{\bs m}(t)\cdot{\bs \sigma}+\Delta_0\tau_x\,.
    \label{HamFull}
\end{align}
The instantaneous spectrum instead consists now of $4$ energy bands $\pm\epsilon_{1,2}(p,t)$, with 
\begin{align}
\epsilon_{1,2}(p,t)&=\sqrt{|{\bs m}(t)|^2+v_F^2p^2+\Delta_0^2\mp2\sqrt{[m_x^2(t)+m_y^2(t)]\Delta_0^2+m_z^2(t)(\Delta_0^2+v_F^2k_x^2)}}\,.
\end{align}
We mention that for $\Delta_0<|{\bs m}(t)|$ and $\theta=0,\pi$ ($m_{x,y}=0$), the spectrum becomes gapless and the adiabatic description is not valid anymore. Thus, we need to exclude from the magnetization dynamics the trajectories that contain these points. We also point out that for general $\theta$ the system undergoes a topological phase transition at $\Delta_0=|{\bs m}|$ so also these points should be avoided.  

While for arbitrary precession angles $\theta$ the analytical expressions for $Q_e$ is in general very long and uninspiring, at $\theta=\pi/2$ we have been able to obtain a simple expression for this quantity:
\begin{align}
    Q_e=e \frac{2r+(1-r^2)\log\left|\displaystyle{\frac{r-1}{r+1}}\right|}{4r}, \ \ \  r\equiv|{\bs m}|/\Delta_0. \label{analytical_charge}
\end{align}
This expression is compared to the numerical calculations in Fig.~\ref{QBerry}.

We see that the charge carried by the pumped quasiparticles is not quantized anymore, which can be interpreted as the superconducting condensate impeding the charge flow. That in turn implies a coupling between the SC condensate and the magnetization dynamics, indicating that the SC condensate can be manipulated by spintronics means.

\section{Instantaneous expectation value of the current}\label{insta_sect}

\begin{figure}
    \centering
    \includegraphics[width=0.5\linewidth]{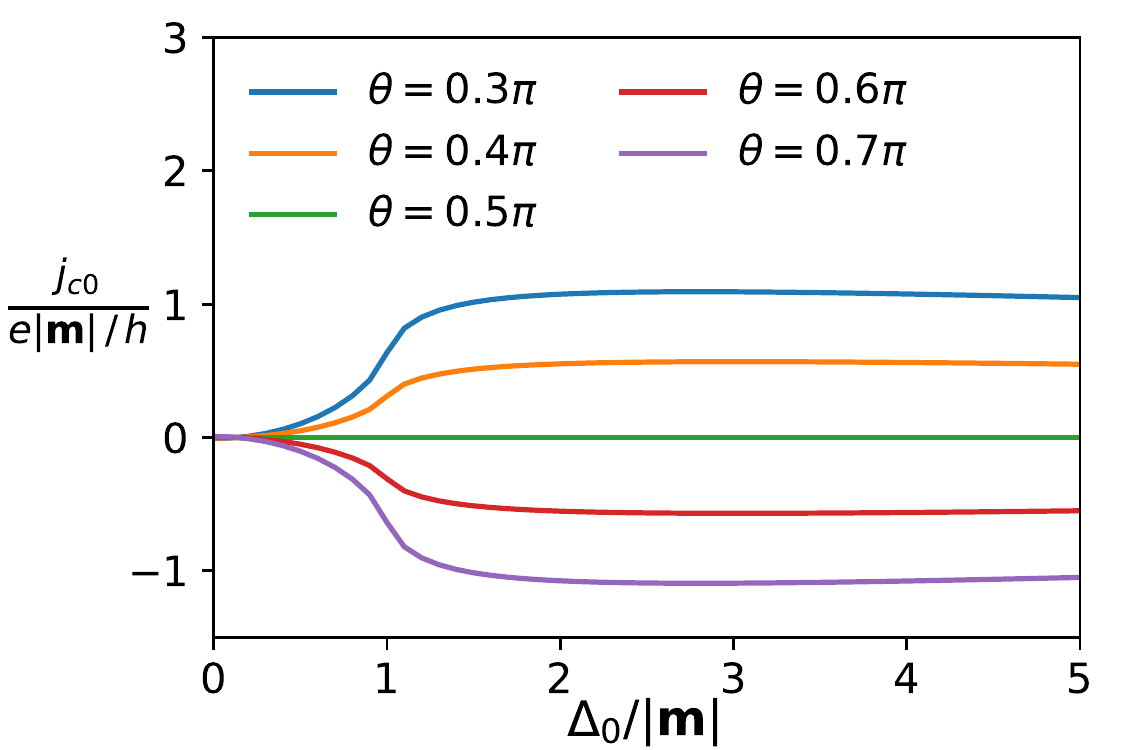}
    \caption{Instantaneous expectation value of the current as a function of $\Delta_0/|{\bs m}|$ for various precession angles. The parameters of the lattice model (\ref{2DtBdGNHam}) are described in Eqs.~(\ref{m0ylattice}-\ref{Deltaylattice}) and in Table \ref{paramtable}.} 
    \label{jinsta}
\end{figure}

Finally we point out that in general the instantaneous expectation value of the current 
\begin{equation}
j_{c0}(\phi)=\frac{e}{2h}\sum_{n\in occ} \int_0^{2\pi} dk~\widetilde{\langle n k, \phi|}\partial_{k}\mathcal{H}_{BdG}(k, t) \tau_z \widetilde{|n k, \phi \rangle}
\end{equation}
is not zero in the presence of superconductivity although the system is gapped. Due to the spin-rotation symmetry around the $z$-axis $j_{c0}(\phi)$ is independent of $\phi$ and it is denoted $j_{c0}$ in the following.

Similarly, as in the case of normal systems the fully occupied bands cannot carry quasiparticle current. However, in the presence of superconductivity the charge current operator is multiplied with $\tau_z$, and therefore $j_{c0}(\phi)$ can be non-zero. This kind of situation occurs for example if the phase-gradient introduces a supercurrent in the system. 

Interestingly, due to the spin-orbit coupling also a Zeeman field ${\bs m}$ can introduce non-zero supercurrent $j_{c0}$. If $m_z=0$ we find that $j_{c0}=0$ because the Hamiltonian satisfies a mirror symmetry. On the other hand, turning on $m_z \ne 0$ we obtain $j_{c0} \ne 0$ and the sign of $j_{c0}$ is determined by the sign of $m_z$ (see Fig.~\ref{jinsta} ). From our simulations we obtain that for $\Delta_0/|{\bs m}| \gg 1$ the instantaneous current is well fitted by the formula $j_{c0}/(e |{\bs m}|/h)= a \cos\theta$, where $a=1.85$. All these features demonstrate that there exists a non-trivial coupling between the superconductivity and magnetism which deserves a more careful study but goes beyond the scope of the present work.